\documentclass[reprint, prl, twocolumn]{revtex4-1}%\usepackage

\usepackage{dcolumn}% Align table columns on decimal point

\usepackage{graphicx}
\usepackage{amssymb,amsmath,amsthm}
\usepackage{epstopdf}
\usepackage{natbib}
\newcommand{\dd}[2]{\frac{d #1}{d #2}}
\newcommand{\pp}[2]{\frac{\partial #1}{\partial #2}}
\newcommand{\R}{\mathbb{R}}
\newcommand{\oneover}[1]{\frac{1}{#1}}

\newcommand{\kb}{k_B}

\newcommand{\grad}{\nabla}

\newcommand{\gradt}[1]{\text{ grad$_{#1}$ }}
\newcommand{\divt}[1]{\text{ div$_{#1}$ }}

\newcommand{\eps}{\epsilon}
\newcommand{\half}{\frac{1}{2}}

\begin{document}

%\conflictofinterest{Conflict of interest footnote placeholder}

%\track{Insert 'This paper was submitted directly to the PNAS office.' when applicable.}

\title{A geometrical approach to computing free energy landscapes from short-ranged potentials}

\author{Miranda Holmes-Cerfon, 
Steven J. Gortler, and  Michael P. Brenner}
\affiliation{
School of Engineering and Applied Sciences and Kavli Institute for Bionano Science and Technology, Harvard University, Cambridge, MA 02138, USA}

\begin{abstract}
Particles interacting with short-ranged potentials have attracted increasing interest, partly for their ability to model mesoscale systems such as colloids interacting via DNA or depletion. 
We consider the free energy landscape of such systems as the range of the potential goes to zero. In this limit, the landscape is entirely defined by geometrical manifolds, plus a single control parameter. These manifolds are fundamental objects that do not depend on the details of the interaction potential, and provide the starting point from which any quantity characterizing the system -- equilibrium or non-equilibrium -- can be computed for arbitrary potentials.
To consider dynamical quantities we compute the asymptotic limit of the Fokker-Planck equation, and show that it becomes restricted to the low-dimensional manifolds connected by ``sticky'' boundary conditions. To illustrate our theory, we compute the low-dimensional manifolds for $n\leq 8$ identical particles, providing a complete description of the lowest-energy parts of the landscape including floppy modes with up to 2 internal degrees of freedom.  The results can be directly tested on colloidal clusters. This limit is a novel approach for understanding energy landscapes, and our hope is that it can also provide insight into finite-range potentials. 
\end{abstract}

\maketitle

The dynamics on free energy landscapes is a ubiquitous paradigm for characterizing molecular and mesoscopic systems, from atomic clusters, to protein folding, to colloidal clusters.  \cite{walesBook, stillinger1984,fah2009,meng2010}.
 The predominant strategy for understanding the dynamics on an energy landscape has focused on the stationary points of the energy, the local minima and the transition states, and seeks the dynamical paths which connect these to each other, while more recent models generalize to metastable states connected by paths as a Markov State Model \cite{bowman2010}.  These techniques have proven to be extremely powerful, giving innumerable insights into the behaviour of complex systems 
 \cite{doyewales1996c, stillinger1995,onuchicwolynes1997, liwo1999,wolynes2005,rothemund2006,clementi2008,maragliano2010},
On the other hand, a major issue has been the difficulty of finding the transition paths, connecting local minima or metastable states to each other, especially given a complex energy landscape in a high-dimensional space.  A variety of creative methods have been developed in recent years for efficiently finding transition paths   
 \cite{swendsen1986,karplus1987,eve2002, chandler2002,hummer2004,abramseve2010, wales1999, wales2005,wales2006,henkelman2000, noe2009}
 but for a given system, there is no guarantee that all such paths have been found.

Here we present a different point of view for understanding an energy landscape, that occurs when the range over which particles interact is much smaller than their size. Such is the case in certain mesoscale systems, for example, for $C_{60}$ molecules \cite{hagen1993, doyewales1996b}, or for colloids interacting via depletion \cite{meng2010} or coated with polymers or complementary DNA strands  \cite{gazzillo2006, dreyfus2009, mirkin2011, crocker2011}. 
We will show that in this limit, the free energy landscape is described entirely by geometry, plus a single control parameter $\kappa$ that is a function of the temperature, depth, and curvature of the original potential.  This limit is related to the sticky sphere limit of a square-well potential \cite{baxter1968}, which has been used to investigate thermodynamic properties of hard sticky spheres \cite{stell1991,frenkel2003,gazzillo2004}.
The landscape can be thought of as a polygon living in a high-dimensional space, whose corners (0-dimensional manifolds) are joined to each other by lines (1-dimensional manifolds), that in turn form the boundaries of faces (2-dimensional manifolds), and so on. These manifolds are fixed functions of the particles' geometries, independent of the details of the original interaction potential from which the limit was taken. 

Once the geometrical manifolds comprising the landscape are computed,  non-equilibrium quantities characterizing the dynamics can be calculated by solving the Fokker-Planck equation or its adjoint on these manifolds. We show that in the short-ranged limit these equations acquire an effective boundary condition at the boundary of every $p$-dimensional manifold in the polygon. This makes the kinetics computationally tractable, since the stiff modes of a narrow potential become a set of boundary conditions. 

The geometrical nature of the energy landscape does not mitigate its high dimensionality, but at low temperatures (high $\kappa$) both the free energies and the kinetics are dominated by the lowest-dimensional manifolds. This means that the description of the free energy landscape and kinetics for short-ranged potentials reduces to a problem in discrete and computational geometry -- to find all of the low dimensional manifolds for a given set of interacting particles.  

As an illustration of the framework, we characterize the geometrical landscape for 
 $n\leq 8$ particles with identical potentials, and demonstrate how these solutions lead to a complete description of the energy landscape and the kinetics of this system.   
This solution describes both the geometrical limit of small atomic clusters as well as a direct prediction for colloidal clusters interacting with depletion forces \cite{doyewales1996c, malins2009, wales2010, meng2010}, where the predictions could be tested experimentally.  The solution also provides a framework for understanding and extending simulations on clusters with finite range potentials \cite{walesBook}.
Our calculation of the energy landscape builds on the enumeration of all finite sphere packings of $n$ particles with at least $3n-6$ contacts \cite{arkus2009,arkus2011}. With these as the starting point, we compute every 1- and 2-dimensional manifold of motions that contains a 0- dimensional manifold at its boundary, from which we can extract statistical quantities such as the relative entropies of the different types of motions. Then, we solve Fokker Planck equations on these manifolds to obtain transition rates between the  lowest energy states, the 0-dimensional manifolds.  

\begin{figure}
\begin{center}
\includegraphics[scale=1]{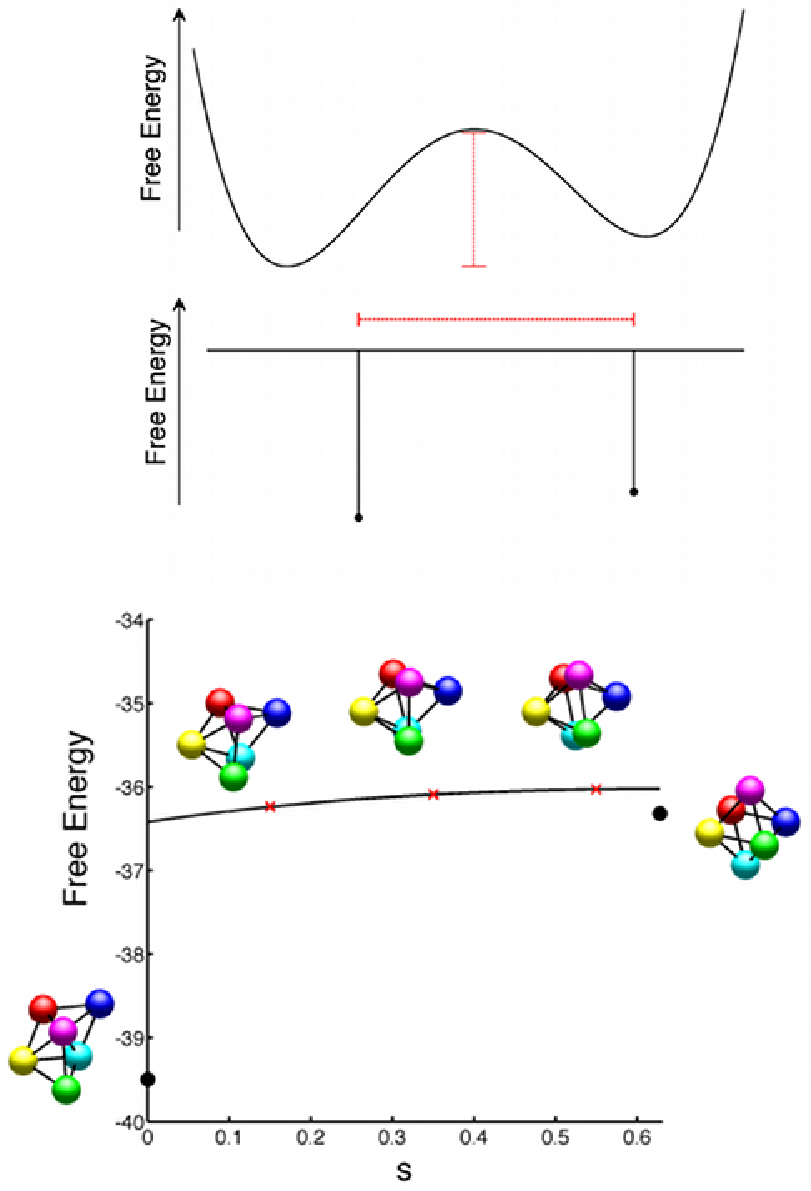}
\end{center}
\caption{Top: schematic of a traditional depiction of an energy landscape. Local minima are separated by energy barriers (red line) that govern the transition rate from one basin to another. 
Middle: schematic of a geometrical energy landscape, showing the 0- and 1--dimensional sets. Local minima are infinitesimally narrow, deep points,  separated by long, nearly flat lines. Along these the dynamics is governed mainly by diffusion, so the length of the line (in red) determines the transition rate. 
Bottom: Example of a 1-dimensional manifold from the landscape for $n=6$, showing the transition path from polytetrahedron (left) to octahedron (right). Black line is the free energy $F_\alpha/\kb T=-\log n_\alpha h_\alpha I - 11\log\kappa$ along the 1-dimensional manifold in units of $\kb T$, where we chose $\kappa=20$. Black dots are the free energy  $F_\alpha/\kb T=-\log n_\alpha h_\alpha I - 12\log\kappa$ for the 0-dimensional endpoints. Red crosses mark locations of clusters that are plotted explicitly (they are plotted with 1/2 their actual diameter for clarity), and
the horizontal axis is a parameterization of the manifold in the quotient-space distance $s$. 
}
\label{fig:schematic}
\end{figure}

\section{The Geometrical Landscape}\label{sec:landscape}

We begin by showing how the geometrical free energy landscape arises as a distinguished limit of particles interacting with arbitrary potentials of  finite range.  We consider 
a point $x\in\R^N$ in configuration space, with the potential energy given as a sum of potentials concentrated near different geometrical boundaries as
\begin{equation}\label{eq:Ueps}
U^\epsilon(x) = \sum_kC_\epsilon U(y_k(x)/\epsilon) .
\end{equation}
The functions $y_k:\R^n\to\R$, ($k=1,\ldots,k_{max}$) represent the geometrical boundaries via their level sets $\{y_k(x)=0\}$, and $U(y):\R\to\R$ is the potential energy near each boundary. 
For concreteness, let us suppose this is a model for $n$ spheres with centers at $x_i\in \R^3$, $i=1\ldots n$, so the configuration is $x= (x_{11}, x_{12}, x_{13}, x_{21}, \ldots, x_{n3}) \in \R^{3n}$,
and we take $y_k$ to be the excess bond distance, as
$y_k = \sqrt{|x_{i(k)}-x_{j(k)}|^2}-d$, where $d$ is the particles' diameter.
Then $U(y)$ is the pairwise interaction potential that we assume has minimum $U_0$ at $y=0$ and  is negligible beyond some cutoff $r_c$.  For ease of exposition, the interaction potential is assumed to be identical between each pair of particles, but this is not a necessary restriction for the geometrical landscape to apply.  
For the total potential $U^\epsilon$, the parameter $\epsilon$ characterizes the range of the potential, while $C_\epsilon$ is proportional to the depth.  The geometrical limit requires taking $C_\epsilon\to\infty$  as $\epsilon \to 0$ in a manner that we specify momentarily. 

We consider particles that evolve according to the overdamped Langevin dynamics \cite{allen} at temperature $T$:
\begin{equation}\label{eq:dx}
\dd{x}{t} = -\oneover{\gamma}\grad U^\epsilon(x)  + \sqrt{2D}\eta(t),
\end{equation}
 where $\gamma$ is the friction coefficient, $D=(\beta\gamma)^{-1}$ is the diffusion coefficient, $\beta = (\kb T)^{-1}$, $k_B$ is the Boltzmann constant,  and $\eta(t)$ is a $3n$-dimensional white noise. The equilibrium probability for this system is  the Gibbs distribution \cite{landau}:
\begin{equation}\label{eq:mueps}
d\rho^\epsilon(x) = (Z^\eps)^{-1}e^{-\beta U^\epsilon(x)}dx .
\end{equation} 
where $Z^\eps$ is the normalizing constant.

The geometrical free energy landscape occurs when the range $\epsilon\to 0$.  
The relationship between the depth and the range is critical to obtaining an interesting limit. If only the range goes to zero, then particles are bonded 
for shorter and shorter times so that in the limit they simply behave like hard spheres. 
On the other hand, if the depth goes to $-\infty$ too quickly, then the particles simply stick together and only unbind on exponentially long timescales \cite{fatkullin10}.  The interesting limit occurs when  particles stick to each other but
unbind on accessible timescales; for this  we must  choose $C_\epsilon$ so that the 
Boltzmann factor for two particles  
to be bonded to approaches a finite, non-zero constant: 
$\kappa=\lim_{\epsilon\to 0}\oneover{d}\int_0^{\epsilon r_c} e^{- C_\epsilon \beta U(x/\epsilon)}dx$, where we define Boltzmann factors non-dimensionally by scaling by the diameter $d$.  Evaluating the integral using Laplace's method then implies 
\begin{equation}\label{eq:k}
\kappa =\lim_{\epsilon\to 0} \frac{\epsilon\; e^{- \beta C_\epsilon U_0}}{d\sqrt{c\beta C_\epsilon U''(0)}}  
\end{equation}
We call the constant $\kappa$ the \emph{sticky parameter}. 
Note that $\kappa$ is a function of \emph{both} the potential \emph{and} the temperature.
The constant $c = 2/\pi$ for hard-spheres.

In the geometrical limit,
 the probability measure $\rho^\epsilon$ becomes concentrated at the exact locations in configuration space where a bond forms, i.e. on the level sets $\{y_k(x)=0\}$ and all possible multi-way intersections. Thus the limiting probability distribution will be a weighted sum of delta functions, each defined on a manifold corresponding to a different set of bond constraints. The weight of the each delta function depends on the number of bonds and a geometrical factor associated with the entropy of the configuration.  
This gives a geometrical picture of the energy landscape:
 When $\kappa$ is large, the occupation probabilities will be dominated by configurations where the number of bonds $m$ is large.  For identical particles, with $n\le 9$, the maximum number of contacts is $m=3n-6$ \cite{arkus2009,arkus2011}: these are rigid structures that have no internal degrees of freedom, so they correspond to 0-dimensional manifolds, or ``points''.  The next lowest configurations in potential energy are manifolds with $m=3n-7$, which are are obtained from rigid structures by breaking one bond. These have one internal degree of freedom so are 1-dimensional manifolds, or ``lines''. The lines form the boundaries of 2-dimensional manifolds, or ``faces'', when another bond is broken, and continuing up in dimensionality we obtain the entire energy landscape as the union of manifolds of different dimensions. 
 
 Figure \ref{fig:schematic} shows a schematic contrasting this limiting case with the traditional picture of an energy landscape. The traditional picture is of an undulating surface, with local minima connected through saddle points, whose heights provide an activation barrier that determines the transition rates between basins. In contrast, in the geometrical limit, the local minima are infinitely narrow and deep, separated by long, relatively flat spaces in between -- the landscape can be thought of as a golf course punctuated with deep trenches very deep holes. Kinetics on this landscape are determined partly by an activation barrier -- the time it takes to climb out of the hole -- and partly by diffusion.
 
The figure also shows an explicit 1-dimensional manifold taken from the landscape for $n=6$, an example we will return to throughout the text. There are two ground states, the polytetrahedron and the octahedron \cite{arkus2011}, 
and the manifold is the set of deformations corresponding to the transition path between these when a single bond is broken.

To explicitly calculate the equilibrium probabilities of the different states in the geometrical landscape we
consider a configuration with $m$ constraints or equivalently  
$p\equiv 3n-6-m$
 bonds broken. The constraints are written as an ordered multiindex $\alpha = (\alpha_1,\alpha_2,\ldots \alpha_m)$ so the manifold of configurations satisfying such constraints is 
\begin{multline}\label{eq:omega}
\Omega_{\alpha} = \{x: y_{\alpha_1}(x) = y_{\alpha_2}(x) = \ldots = y_{\alpha_m}(x) = 0,\\
y_\beta(x)>0, \beta \neq \alpha_i \}.
\end{multline} 
We write $\alpha=0$ to mean the region where no constraints are active, and let $\Omega = \cup_{\alpha} \Omega_\alpha$ be the full space of accessible configurations. 
 The limiting partition function associated with these constraints is
\begin{equation}\label{eq:lim}
z_\alpha = \lim_{\epsilon\to 0} \oneover{d^{3n}}\int_{\Omega_\alpha^\epsilon}  e^{-\beta U^\eps(x)}dx ,
\end{equation}
where $\Omega_\alpha^\epsilon$ is the neighbourhood surrounding the manifold where the potential $U^\eps$ associated with  the constraints is active:
\begin{multline}\label{eq:omegaeps}
\Omega_{\alpha}^\eps = \{x: 0\leq y_{\alpha_1}(x),\ldots,y_{\alpha_m}(x)\leq \eps r_c,\\
y_\beta(x)> \eps r_c, \beta \neq \alpha_i \}.
\end{multline} 
This splits configuration space near each manifold into two parts -- the fast variables $y_{\alpha_i}$ changing rapidly along directions associated with the constraints (sometimes called the vibrational modes), and the slow variables $y_\beta$ that are the unconstrained configuration.

To compute the integral in \eqref{eq:lim} we need a parameterization of the manifolds associated with the constraints. 
It is convenient to parameterize the fast directions by the constraint variables themselves,
 $x\to y_{\alpha_1}, \ldots y_{\alpha_m}$.  We choose  the additional $3n-m$ variables $y\in\R^{3n-m}$ so that 
$\grad y \cdot \grad y_{\alpha_i} = 0$ on $\Omega_\alpha$, i.e. the variables $y$, $y_{\alpha_i}$ are orthogonal on the manifold.
As discussed in the Appendix, it is possible to find such a parameterization locally as long as the manifold associated with the constraint variables is regular -- i.e., the coordinate transformation for the constraint variables must be smooth and invertible. This happens when 
the Hessian of the potential energy (say at $\eps=1$) 
\begin{equation}\label{eq:H}
H_\alpha(x) =  \grad\grad U^{\eps=1}(x) = U''(0)\sum_{i=1}^m \grad y_{\alpha_i}(\grad y_{\alpha_i})^T 
\end{equation} has $m$ non-zero eigenvalues.  
 
We can now evaluate
\begin{equation}\label{eq:int1}
z_\alpha = \oneover{d^{3n}}\int_{\Omega^\eps_\alpha} \epsilon^m e^{-\beta C_\eps \sum_{i=1}^mU(Y_i)}\sqrt{|g_{ij}|} dYdy
\end{equation}
where $Y = (Y_1,\ldots, Y_m)$ with $Y_i = y_{\alpha_i}/\eps$, $g_{ij}$ is the metric tensor associated with the transformation $(y_{\alpha_i},y)\to x$ with components $g_{ij} = J_{ki}J_{kj} = J^TJ$, where  $J_{ij} = \pp{x_i}{y_j}$, and $|g_{ij}|$ is its determinant. Let separate the metric tensor into blocks as
\begin{equation}\label{eq:gblock}
g =  \left( \begin{array}{cc} g_{ab} & g_{av} \\ g_{ub} & g_{uv} \end{array}\right) .
\end{equation}
The first set of indices run from $1\leq a,b \leq m$ and describe the fast variables, perpendicular to the manifold,  while the second set run from $m+1\leq u,v\leq N$ and describe the slow variables, along the constraint manifold. Let $|g_{cd}|$ be the determinant of a particular block.
It follows from the definition of the metric that $|g_{ab}|_{Y=0} = \prod_{i=1}^m \lambda_i^{-1}$, where the $\lambda_i$'s are the non-zero eigenvalues of $H_\alpha(x) / U''(0)$, and the condition of orthogonality gives $|g_{av}|_{Y=0} = |g_{ub}|_{Y=0} = 0$. Evaluating the integral in \eqref{eq:int1} over the fast variables using Laplace's method, and letting $\epsilon\to 0$, shows the limit is
 \begin{equation}\label{eq:int2}
z_\alpha = \kappa^m \oneover{d^{p+6}}\int_{\Omega_\alpha }  h_\alpha(y) \sqrt{|g_{\alpha}|} dy,
\end{equation}
where 
\begin{equation}
h_\alpha(y) = \prod_{i=1}^m \lambda_i^{-1/2}
\end{equation}
is a geometrical factor  (representing the ``vibrational'' degrees of freedom) that depends only on the level sets of the constraints, and 
 $g_\alpha = g_{uv}|_{Y=0}$ is the metric on manifold $\Omega_\alpha$ inherited from the ambient space by restriction. The integral \eqref{eq:int2} is simply the volume integral of $h_\alpha(y)$ over $\Omega_\alpha$. 
  
The manifold $\Omega_\alpha$ contains 6 degrees of freedom representing translation and rotation of the cluster, and the partition function integral can be further simplified by 
 integrating over the subspace spanned by these motions. 
This introduces a factor $I(x)$ in the integral, the square root of the moment of inertia tensor \cite{landau}. If we let $\Omega^Q_\alpha$ be the quotient space formed by identifying points $x\sim z$ if $x$ can be mapped to $z$ by a combination of translations and rotations, i.e. $\Omega^Q_\alpha = \Omega_\alpha / SE(3)$, where $SE(3)$ is the special Euclidean group, then we can write 
\begin{equation}\label{eq:Za}
z_\alpha = \kappa^m\zeta_\alpha=\kappa^m \oneover{d^{p}}\int_{\Omega_\alpha^Q}  h_\alpha(x)I(x) \sqrt{|\bar{g}_\alpha|}dx
\end{equation}
where $\bar{g}_\alpha$ is the metric on $\Omega^Q_\alpha$, and we have dropped constants (such as free volume) that are the same for all configurations.  For convenience later on,
 we define $\zeta_\alpha$ as the part of the partition function that is independent of $\kappa$. The Appendix has a detailed discussion for how to construct $\bar{g}_\alpha$, which is critical for using the formalism developed here for practical calculations.  Fig.
\ref{fig:schematic} (bottom) is an example of such a quotient manifold, where  each point on the manifold represents the 6-dimensional space of clusters in configuration space that are related by rotations or translations. 

The particles in Figure \ref{fig:schematic} are different colours  to identify the different transitions that occur when moving around configuration space. However, when (as imagined here) the particles are identical, permuting the colors of any particular structure yields a geometrically isomorphic structure on a separate part of the quotient space. The free energy  must account for the number $n_\alpha$ of distinct manifolds that are geometrically isomorphic to $\Omega_\alpha^Q$. 
When $p=0$ this is $n_\alpha = C_0N!/\sigma$, where $\sigma$ is the symmetry number, i.e. the number of particle permutations that are equivalent to a rotation, and $C_0=2$  if the structure is chiral and $C_0= 1$ otherwise  \cite{calvo2012}. 
For $p>0$, we count the multiplicities by counting how many times a mode occurs from the perspective of each 0-dimensional ``corner'' of the mode, and dividing by the total number of corners. \footnote{This is a combinatorial argument; it is equivalent to considering the molecular symmetry group for nonrigid molecules as in \cite{walesBook}, Section 3.4.  }
For example, Fig. \ref{fig:schematic} has corners from two different ground states, the polytetrahedron and the octahedron, which occur with multiplicities $n_1, n_2$ respectively. For each polytetrahedron, there is $\nu_1=1$ line coming out of it that is isomorphic to this transition, and for the octahedron there are $\nu_2 = 12$ distinct lines. (The numbers $\nu_1$, $\nu_2$ are indicated on the arrows connecting red circles to blue circles in Figure \ref{fig:landscape6}, where the transition under consideration is mode 7.) 
Therefore the total multiplicity of the line is $(n_1\nu_1+n_2\nu_2)/2$. 
Consider a transition connecting a polytetrahedron to a distinct copy of itself, say mode 5. Here there are $\nu_1 = 4$ such lines connected to each polytetrahedron, so the multiplicity of the line is $n_1\nu_1 / 2$.  
In general, each
 $p$-dimensional manifold $\Omega_\alpha$ contains a total of $n_c$ corners from $N\leq n_c$ nonisomorphic ground states, each ground state having multiplicity $n_i$, and such that each single ground state is connected to $\nu_i$ distinct manifolds isomorphic to $\Omega_\alpha$, so the multiplicity is $n_\alpha = \sum_{i=1}^N n_i\nu_i/n_c$. 

Putting this all together, the total partition function of all structures isomorphic  to a given constraint manifold $\Omega_\alpha^Q$ is $n_\alpha z_\alpha$, and the free energy of these structures are $F_\alpha=-k_B T \log(n_\alpha z_\alpha)$.  We can separate this free energy into $F_\alpha=-m \kb T \log\kappa - \kb T\log(n_\alpha\zeta_\alpha)$, using the definition of $\zeta_\alpha$ in Equation \eqref{eq:Za}.  The first term ($-m\kb T \log\kappa$) depends on the temperature, bond energy, and width of the potential, whereas the second term with  $S_\alpha=- \kb \log(n_\alpha\zeta_\alpha)$ is entirely geometrical, and essentially is the entropy of structures corresponding to the constraint set $\Omega_\alpha^Q$.
As an example,  
we have plotted $F_\alpha/\kb T$ along the polytetrahedral-octahedral transition in  Figure \ref{fig:schematic} (bottom). This varies smoothly along the manifold as $I(x)$, $h(x)$ vary. The endpoints of the manifold are the ground states, where the free energy changes discontinuously because the number of bonds has changed -- this causes a jump in the energetic part via $m$, and the entropic part via $h_\alpha$.

With these results in hand, we can now compare  the total entropies of floppy manifolds of different dimensions, to understand the temperature range in which the different manifolds occur. Let the total geometrical partition function of manifolds of dimension $p$ be
\begin{equation}\label{part}
Z_p = \sum_{\text{dim}(\Omega^Q_\alpha) = p} n_\alpha \zeta_\alpha,
\qquad Z = \sum_p \kappa^{3n-6-p}Z_p .
\end{equation}
Here $Z_p$ is independent of the temperature and potential, while $Z$ combines everything to obtain the entire landscape. Note that  lower dimensional manifolds have more bonds and thus are favored in the partition function when the temperature (or equivalently $\kappa$) is small.
%owing to the temperature dependence of $\kappa$.  
As temperature increases, $\kappa$ shrinks and higher dimensional manifolds become more highly populated. Eventually clusters will fall apart into single particles.   The temperature dependence of how clusters fall apart is encoded in the relative sizes of the $Z_p$'s.
The temperature where the landscape transitions from having more $p$ dimensional structures than $p+1$ dimensional structures is  found by solving $\kappa Z_p = Z_{p+1}$ for $\kappa$, which gives roughly $(\kb T_p)^{-1} \approx \ln Z_{p+1}/Z_p +const$.

\subsection{Free energy landscape for identical particles}

\begin{table*}[ht]
\caption{ Geometric partition functions $Z_p$, and numbers of different modes, for the set of $p=0,1,2$-dimensional manifolds as the number of spheres $n$ varies. These are geometrical quantities that do not depend on the temperature or potential. The total partition function of the $p$-dimensional manifolds, which includes both of these dependencies via $\kappa$, is $\kappa^mZ_p$. }
\label{tbl:freeEn}
\begin{tabular*}{\hsize}{@{\extracolsep{\fill}}l  ccc  ccc  cc}
 $n$ & \# 0-D & \# 1-D & \# 2-D & $Z_0$ & $Z_1$ & $Z_2$ & $Z_1/Z_0$ & $Z_2/Z_1$ \\
 \hline
5 & 1 & 2 & 4 & 10.7 & 73.8 & 545 & 6.9 & 7.4 \\
6 & 2 & 5 & 13 & 36.1 & 256 & 1140 & 7.1 & 4.5  \\
7 & 5 & 16 & 51 & 1.1$\times10^{3}$ & 8.5$\times10^{3}$ & $39\times10^{3}$ & 7.6 & 4.6 \\
8 & 13 & 75 & 281 & 49$\times10^{3}$ & 396$\times10^{3}$ & 1.87$\times10^{6}$ & 8.1 & 4.7\\
\hline
\end{tabular*}
\end{table*}

\begin{figure*}
\includegraphics[width=1\textwidth]{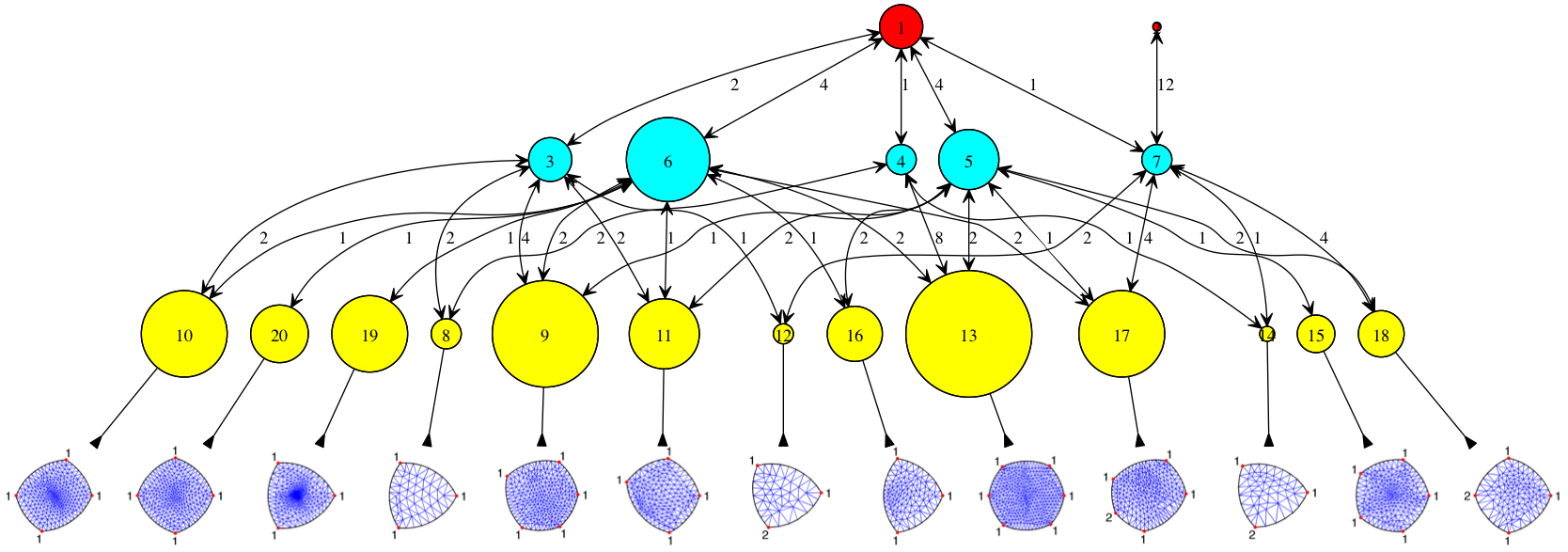}
%\centerline{\includegraphics[width=1\textwidth]{energy_n6_polys_ft}}
\caption{Free energy landscape for n=6 with 0, 1, and 2 bonds broken.
Red circles are 0-dimensional modes, blue are 1-dimensional, yellow are 2-dimensional.
The area of each circle is proportional to the geometrical partition function $n_\alpha\zeta_\alpha$ of each mode, hence to its probability in equilibrium relative to modes of the same dimension.
Modes are identified by numbers and arrows show the connectivity: an arrow from mode $i$ to mode $j$ indicates that mode $i$ is part of the boundary of mode $j$. The number on each arrow indicates the number of different manifolds of type $j$ that are connected to a single manifold of type $i$. 
The computed parameterizations are shown for each of the 2-dimensional modes.
}
\label{fig:landscape6}
\end{figure*}

\begin{figure}
\centerline{\includegraphics[width=1\linewidth]{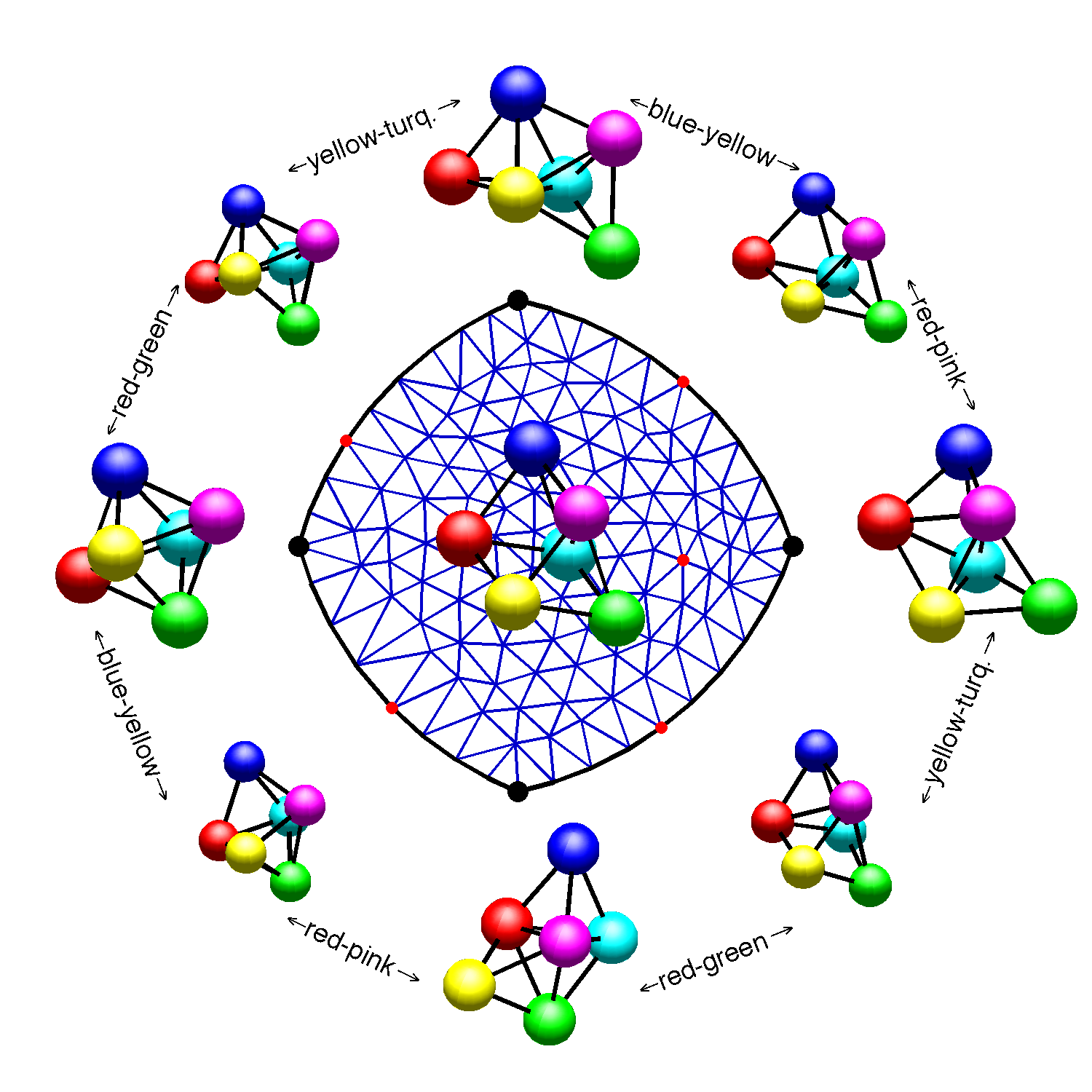}}
%\centerline{\includegraphics[width=0.5\textwidth]{manif18color_arrow}}
\caption{A two-dimensional manifold (mode 18, $n=6$), parameterized in the plane, with selected points on the manifold plotted as clusters. 
The vertex of each triangle represents a cluster and black or red dots indicate the ones that are plotted. The corners (black dots) are rigid structures, or 0-dimensional manifolds. The edges are 1-dimensional manifolds and points on these are obtained from rigid structures by breaking one bond, while  points in the interior form a 2-dimensional manifold and are obtained by breaking two bonds. The 1-dimensional manifolds, beginning at the octahedron (left) and moving clockwise, are 7,5,5,7. 
The text indicates the type of bond that breaks or forms as one moves along each edge.}
\label{fig:2dmanif}
\end{figure}

\begin{figure*}
\includegraphics[width=1\linewidth]{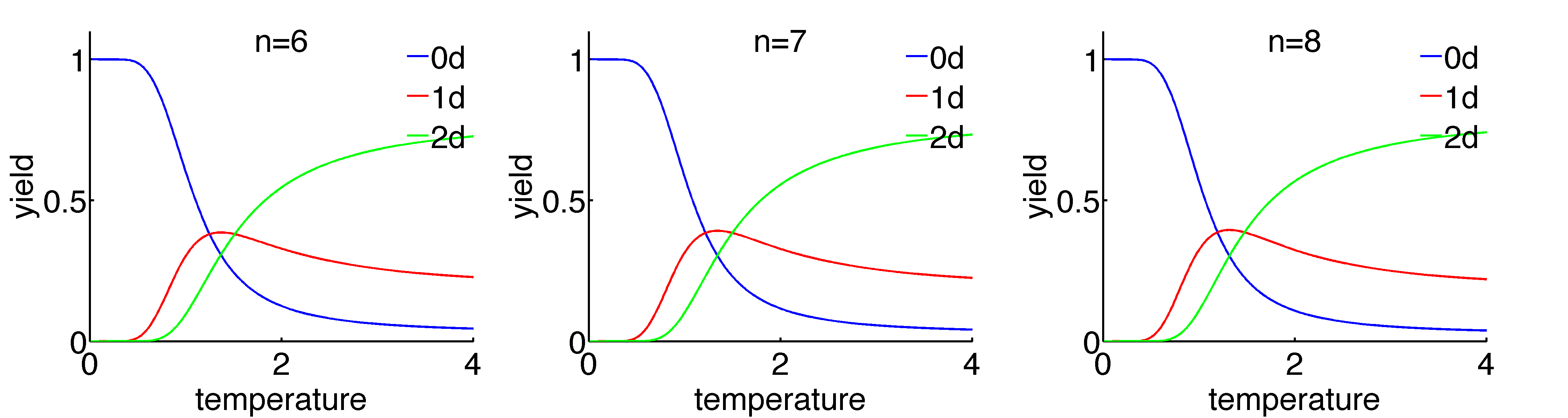}
\caption{Relative yields of 0-dimensional, 1-dimensional, 2-dimenisional modes (neglecting all higher-dimensional modes) for $n=6,7,8$. 
The yield for $p$-dimensional modes is calculated as $y_p = \kappa^{2-p} Z_p / (\kappa^2 Z_0 + \kappa Z_1 + Z_2)$ with $\kappa(T) = e^{-(U_0/\kb)/T}/\sqrt{(U''(0)/\kb) (2/\pi)/T}$. We used $U_0/\kb = -4$, $(U''(0)/\kb) (2/\pi) =15$, but the numbers don't change the qualitative shape.
Note that modes with dimensions $>2$ will become important at the higher temperatures.
}\label{fig:yields}
\end{figure*}

To illustrate our asymptotic calculations with a concrete example we 
have computed the geometrical manifolds up to dimension $p=2$ for 
 $n=5,6,7,8$ identical particles with diameter $d=1$. 
 To do this, we
 begin the set of  clusters with $\geq 3n-6$ bonds  derived in Arkus et. al. \cite{arkus2011} For every rigid cluster we break each single bond in turn and move along the internal degree of freedom until we form another bond. This is the set of one-dimensional manifolds. For the two dimensional manifolds, we break each pair of bonds from the rigid clusters in turn, and move along the internal degrees of freedom to compute the boundaries, corners, and interior of the two-dimensional manifolds (see Appendix for details.) 
This algorithm ensures we have every floppy manifold that can eventually access one of the rigid clusters in our list only by forming bonds, but never breaking them. 
Our analysis makes three assumptions that we believe to be true, but await rigorous proof: First, we are assuming that the list of clusters in Arkus et. al. \cite{arkus2011}  is the complete set of rigid (0-dimensional) clusters; this is true as long as there are no rigid clusters with $3n-7$ bonds or less, a condition that was not checked \footnote{ The existence of such pathological examples is not ruled out purely by rank constraints on the Jacobian as there could be singular structures, see e.g. the examples in Asimow \& Roth (1978) ``The rigidity of graphs'', \emph{Trans. Amer. Math. Soc.} 245:279--289.
}.   Secondly, in the calculations of the entropy of the two dimensional floppy manifolds we assume that the manifolds are
 topologically equivalent to a disk. The fact that our parameterization algorithm works is evidence for this claim, though we have not proved this rigorously. 
 Thirdly, we assume that all floppy manifolds can eventually access a rigid mode and are not, for example, circles. 
 We mention these caveats because although we are confident that they do not apply in the low $n$ examples described here, it is possible that  potentially significant exceptions arise at higher $n$.

 The landscape for $n=6$ is shown \footnote{ Quantitative summaries  are given in Table \ref{tbl:data} and Figure \ref{fig:clustersn6} in the Appendix.}
 in Figure \ref{fig:landscape6}. 
There are two ground states, denoted by the red circles, each with $3n-6=12$ contacts; the area of the red circles are proportional to the probability of each state, with the polytetrahedral ground state $\approx 25$ times more likely than the octahedral. The light blue circles denote the 5 topologically unique structures that are missing a single bond in the ground states -- such structures correspond to a one dimensional manifold, with continuous deformation along the direction of the missing bonds. The yellow circles denote the 13 unique structures that are missing two bonds from the ground state.  Again, the area of the circles is proportional to the occurrence probability of these modes.
 These structures correspond to two dimensional manifolds, with continuous deformations allowed along both of the directions between the missing bonds.
  The connections between the different
 modes are denoted by arrows on this figure, with structures missing (say) 2 bonds generally arising from breaking a single bond from several  different pathways.

Each element of a mode with two bonds broken can be mapped to a polygon in $\R^2$, and these parameterizations are also shown in Figure \ref{fig:landscape6}. We have chosen one parameterization (mode 18), to illustrate in detail in 
Figure \ref{fig:2dmanif}. The interior of the manifold represents structures with 10 bonds, with each point representing a different set of coordinates for the particles.  The edges correspond to structures with 11 bonds, while the corners are structures with a full $3n-6=12$ bonds. Of the four corners, three correspond to  polytetrahedra with different permutations of the particles, and the final one is an octahedron. The one-dimensional edges connecting the corners are possible transition paths that can be followed by breaking only one bond. 

In general the number of corners varies among modes, with no a priori way to determine this without solving the full geometry problem. It 
ranges from 3--6 for $n=6,7$, and from 3--7 for $n=8$. Many 2-dimensional modes contain several permutations of a given 0-dimensional mode as a corner.

Table 1 summarizes the partition function data. 
The number of different modes increases combinatorially with $n$, as do the geometrical partition functions $Z_p$. Strikingly,
 the ratios $Z_1/Z_0$, $Z_2/Z_1$ remain virtually constant as $n$ increases. 
 This implies that the temperature dependence of the landscape is independent of $n$. 
 Indeed, Figure \ref{fig:yields} shows the relative probabilities of 0,1,2-dimensional modes as the temperature varies, for $n=6,7,8$, with the yield of a $p$-dimensional mode given by $y_p = \kappa^{2-p} Z_p / (\kappa^2 Z_0 + \kappa Z_1 + Z_2)$.  As an illustration, we have chosen $U_0/\kb = -4$, $(U''(0)/\kb) (2/\pi) =15$, so that
  $\kappa(T) = e^{4/T}/\sqrt{15/T}$.  
 Because the ratios $Z_{p+1}/Z_p$ are nearly the same, these graphs are essentially indistinguishable for different numbers of particles.  Moreover the critical temperature for transitioning from mostly $0$ dimensional structures to $1$ dimensional structures ($\approx 1/\ln Z_1/Z_0$) is quite close to that for transitioning from 1 dimensional structures to 2 dimensional structures ($\approx 1/\ln Z_2/Z_1$).   If this 
remains true as $p$ increases, it would imply that clusters melt explosively at some critical temperature, rather than incrementally:
clusters would occupy either mostly the 0-dimensional modes, or a gaseous, no- or few-bond-state, but not the chain-like floppy configurations in between.

\section{Kinetics on the geometrical landscape}

\begin{figure}
\includegraphics[width=1\linewidth]{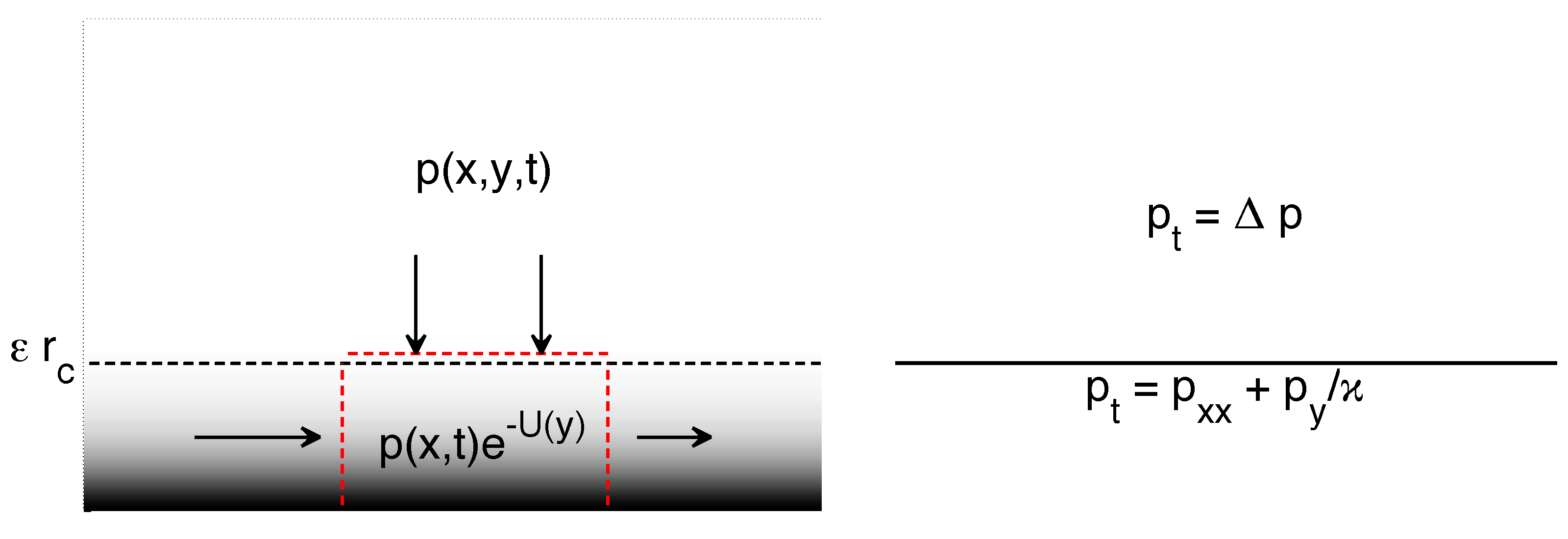}
%\begin{minipage}[c]{0.25\linewidth}
%\includegraphics[width=1\linewidth]{plim-before-1d}
%\end{minipage}\hfill
%\begin{minipage}[c]{0.25\linewidth}\begin{center}
%\includegraphics[width=1\linewidth]{plim-after-1d}\\
%\end{center}
%\end{minipage}
\caption{Schematic showing the asymptotic boundary condition near a one-dimensional boundary. On the left, grey shading indicates the depth of the potential, and dashed line indicates the boundary at which the outer and inner solutions. By considering the total probability flux in/out of each small volume element near the boundary (red box), one can replace the detailed dynamics near the boundary in the limit as $\epsilon r_c\to 0$ with an effective boundary condition (right). 
}
\label{fig:plim}
\end{figure}

We now consider kinetics on the geometrical landscape.
The concentration of  equilibrium probabilities  on manifolds with varying dimensions also applies
to non-equilibrium quantities, such as transition rates, first-passage times, the evolution of probability density, etc.    Such quantities 
 can be computed from the time-dependent transition probabilities, which for dynamics given by \eqref{eq:dx} are obtained from the corresponding forward or backward Fokker-Planck (FP) equations \cite{gardiner}. We show that in the geometrical limit, the FP equation asymptotically approaches a hierarchy of FP equations, one on each manifold of each dimension. These equations are coupled to each other, with equations on manifolds with dimension $p$ serving as boundary conditions for the equations on manifolds with dimension $p+1$. 

The idea behind the derivation is quite natural: if a potential is deep and narrow, then it equilibrates much more rapidly in the directions along the bonds than along a cluster's internal degrees of freedom. Therefore the probability density near a cluster with $p$ bonds broken is approximately $p(y,t)e^{-\beta\sum_i U^\epsilon (Y_{\alpha_i})}$, where $y$ parameterizes the internal degrees of freedom of the cluster. The ``constant'' $p(y,t)$ evolves slowly in the transverse directions according to the Fokker-Planck dynamics (which includes an ``effective'' potential that arises from the curvature of the manifold), and it also changes due to the flux of probability out of the $p+1$-dimensional manifolds for which it forms part of their boundary. This gives a hierarchy of coupled FP equations. 

To see how this comes about in detail, we examine solutions to the Fokker-Plank equation for the evolution of the probability density $p^\eps(x,t)$. Given a parameterization of configuration space $\R^{3n}$ with metric tensor $g$, the non-dimensional Fokker Plank equation corresponding to \eqref{eq:dx}  is 
\begin{multline}\label{eq:fp}
\partial_t p^\eps = \divt{} ( p^\eps \gradt{} U^\epsilon + \gradt{} p^\eps) \\
= \oneover{\sqrt{|g|}} \partial_i\left( \sqrt{|g|} \left( p^\eps g^{ij}\partial_j  U^\epsilon + g^{ij}\partial_j p^\eps\right)\right), 
\end{multline}
with boundary conditions at each level set $\{y_i(x)=0\}$
\[
(p^\eps \gradt{}  U^\epsilon + \gradt{} p^\eps)\cdot \hat{n}^{(i)} = \left( p^\eps g^{ij}\partial_j  U^\epsilon + g^{ij}\partial_j p\right) \hat{n}^{(i)}_j = 0 
\]
where $\hat{n}^{(i)}$ is the outward normal to the boundary. We have non-dimensionalized lengths by $d$,  times by $d^2/D$, and energy by $\kb T$.  
Away from all boundaries there is no force, so in $\Omega_0$ the limiting probability $p$ evolves only by diffusion as
\begin{equation}\label{eq:p0}
\partial_t p = -\text{ div } j_0, \qquad \text{with}\quad j_0 = -\text{ grad } p.
\end{equation}

Now consider the evolution near a manifold $\Omega_\alpha$, a ``wall''. 
The dynamics in the directions orthogonal to the wall, where bond distances are changing, are much faster than those along it, so near the wall the probability density will rapidly approach a multiple of the equilibrium probability. 
Parameterizing the region $\Omega^\epsilon_\alpha$ near the wall  as $(y_{\alpha_i},y)$ as in the previous section, we obtain
\begin{equation}\label{eq:p1s}
p^\epsilon(y_{\alpha_i},y,t) = p(0,y,t)e^{- \sum_{i=1}^m U^\epsilon(y_{\alpha_i})} + \epsilon p_1e^{- \sum_{i=1}^m U^\epsilon(y_{\alpha_i})} ,  
\end{equation} 
where $\epsilon p_1(y,y_{\alpha_i},t)$ is the correction to the leading order formula. This satisfies the matching condition that $p^\eps$ be asymptotically continuous. 
This ansatz can also be derived from a consistent asymptotic expansion of \eqref{eq:fp} after the change of variables $Y_{\alpha_i} = y_{\alpha_i}/\epsilon$.

Substituting \eqref{eq:p1s} into  the Fokker-Planck equation \eqref{eq:fp}  gives 
\begin{equation}\label{eq:fp2}
\partial_t\left(  e^{-U^\epsilon} p  \right) 
= \divt{} \left(  e^{- U^\epsilon} \gradt{} p  \right) + O(\epsilon).
\end{equation}
We would like to to  integrate out the fast variables so we need to separate these from the slow ones. This is most conveniently done using the metric $g$ with block decomposition \eqref{eq:gblock}. Making the change of variables $Y_i = y_{\alpha_i}/\epsilon$ shows that \eqref{eq:fp2} can be written as
\begin{multline}\label{eq:fp3}
\partial_t\left(  \sqrt{|g|}e^{- U_\alpha} (p+p_1)  \right) 
= \oneover{\epsilon^2}\partial_a \left(  \sqrt{|g|} e^{-  U_\alpha} g^{ab}\partial_b (p+\epsilon p_1)   \right)  \\
+ \oneover{\epsilon}\partial_a \left(  \sqrt{|g|} e^{-  U_\alpha} g^{av}\partial_v (p+\epsilon p_1)   \right)\\
+ \oneover{\epsilon}\partial_u \left(  \sqrt{|g|} e^{- U_\alpha} g^{ub}\partial_b (p+\epsilon p_1)   \right)\\
+ \partial_u \left(  \sqrt{|g|} e^{- U_\alpha} g^{uv}\partial_v (p+\epsilon p_1)   \right)
.\phantom{+O(\epsilon)} 
\end{multline}
where we abbreviate $U_\alpha = \sum_{i=1}^m C_\epsilon U(Y_i)$. 

We now integrate \eqref{eq:fp3} over the fast variables $\eps^m dY_1dY_2\ldots dY_m$ and keep the leading-order parts.
The terms $\propto\oneover{\epsilon}$ vanish in the limit because $g^{av}|_{Y=0}= g^{ub}|_{Y=0}=0$. 
The $O(1)$ terms require evaluating an integral similar to \eqref{eq:int1} which converts $\sqrt{|g|} e^{- U_\alpha}$ to the factor $\sqrt{|g_{uv}|}\kappa^mh_{\alpha}$ at each point on the manifold.

The first term is the most interesting. This is the divergence in the fast variables, and although $p$ does not depend on these, the unknown $p_1$ might and contributes at $O(\epsilon^{-1})$. However, we can use the divergence theorem to replace this term with an integral of the flux through the $p$-dimensional fast-variable boundary, at $\{x:Y_i(x)=r_c\}$. We then introduce a second matching condition which requires the flux to be continuous, so we can replace it with the sum of fluxes from each $p+1$-dimensional manifold $\Omega_\beta$ which has $\Omega_\alpha$ as part of its boundary, 
\footnote{Specifically, we require that
$
\epsilon \tilde{j}_\alpha\cdot  \hat{n}^{r_c}|_{Y_i = r_c} = \tilde{j}_{\beta}\cdot \hat{n}^{(i)}|_{y_{\alpha_i}=0},
$
where $\hat{n}^{r_c}$ is the normal to level set $Y_i=r_c$ and $\hat{n}^{(i)}$ is the normal to level set $y_i=0$, and $\tilde{j}_\alpha = - e^{- U_\alpha} g^{ab}\partial_b (p+\epsilon p_1) $ with a similar expression for $\tilde{j}_\beta$.  
}
and evaluate the limit of the boundary integral for these matched fluxes.

Finally, we integrate over the space of rotations and translations of each point, assuming $p(x,0)$ is constant on orbits. The divergence in these directions will disappear by Stokes' theorem, and the remaining directions provide dynamics on the quotient space.
Combining with the previous calculations, yields 
\begin{multline}\label{eq:SFP}
\partial_t(\kappa_\alpha p) = \oneover{\sqrt{|\bar{g}_{uv}|}}\partial_u\left( \sqrt{|\bar{g}_{uv}|} \kappa_\alpha \bar{g}^{uv} \partial_v p\right) + \sum_{\beta\to\alpha}j_\beta\cdot\hat{n}^{\beta\alpha} \\
= \divt{\alpha}(\kappa_\alpha \gradt{\alpha} p) + \sum_{\beta\to\alpha}j_\beta\cdot \hat{n}^{\beta\alpha}
\end{multline}
where the fluxes have leading order part
\begin{equation}
 j_\alpha =  - \kappa_\alpha \gradt{\alpha} p \qquad \text{on }\Omega^Q_\alpha.
\end{equation}
Here $\kappa_\alpha \equiv \kappa^m h_\alpha I$ combines the sticky parameter and the geometric factor at the wall, and the metric tensor $\bar{g}$ is the quotient metric obtained from the metric $g_\alpha$ on $\Omega^Q_\alpha$, which in turn is inherited from the original metric $g$ in the ambient space by restriction: $g_\alpha = g|_{Y_1=\ldots=Y_m=0}$. 
The sum is over $\beta$ such that $\Omega^Q_\alpha$ is part of the boundary of the $p+1$-dimensional manifold $\Omega^Q_\beta$, and where $\hat{n}^{\beta\alpha}$ is an outward normal vector to $\Omega^Q_\beta$ at $\Omega^Q_\alpha$. We define  $\divt{\alpha}, \gradt{\alpha}$ to be the differential operators on the quotient manifold $\Omega_\alpha^Q$. 

Equation 
 \eqref{eq:SFP} has a more intuitive interpretation as the evolution of the probability \emph{along the wall}. It is clear from the derivation that the total probability density (with respect to the wall coordinates) of being on the wall is $P_\alpha \equiv \kappa_\alpha p$. This satisfies
\begin{multline}
\partial_t P_\alpha = \\
   \oneover{\sqrt{|\bar{g}_{uv}|}}\partial_u \left( \sqrt{|\bar{g}_{uv}|} \left( -P_\alpha \bar{g}^{uv}\partial_v\log\kappa_\alpha + \bar{g}^{uv}\partial_v P_\alpha \right)\right) +  j_\beta\cdot \hat{n}^{\beta\alpha}\\
= \divt{\alpha} \left(-P_\alpha\underbrace{\gradt{\alpha} \log\kappa_\alpha}_{\text{effective force}} + \underbrace{\gradt{\alpha}P_\alpha}_{\text{diffusion}} \right)  + \underbrace{j_\beta\cdot \hat{n}^{\beta\alpha} }_{\text{flux to/from wall}} . \label{eq:P1}
\end{multline}
The dynamics along the wall is therefore a combination of diffusion, plus drift due to an ``effective'' potential $-\log \kappa_\alpha$, plus a flux in and out of the wall.  

The effective potential is entropic in nature and comes from the changing wall curvature, which makes the potential look wider in some places than others. A particle will spend more time in the wide places than in the narrow ones, and since it reaches equilibrium much more quickly in the transverse directions than in the along-wall directions, it looks like there is an effective force pushing it to the wider areas.
%-- see Figure \ref{fig:parabola} for a simple illustration. 
This is the same equation one obtains by letting the depth of the potential become infinite without changing the width \cite{fatkullin10}, but with the addition of the flux in and out of the wall.

This flux term is illustrated schematically in Figure \ref{fig:plim} for a simple case where the configuration space $(x,y)\in\R^2$ has a single constraint, $y_1(x,y) = x$ -- a  ``wall'' at the horizontal axis. 
The probability integrated over a small box (red) with length $\Delta x$ near the wall changes in a time increment $\Delta t$  due to two processes: probability fluxing along the wall in the $x$-coordinate, which contributes a change of $\Delta t (p_x(x+\Delta x,t)-p_x(x,t))\int e^{- U(y)}dy$, and the flux of probability from the wall to the interior, which contributes a change of $\Delta t \Delta x p_y(x,0,t)$. Equating with $\Delta t p_t(x,t)\int e^{-U(y)}dy$ gives the effective boundary condition.

To summarize, the limiting FP equation and boundary conditions are \eqref{eq:p0}, plus \eqref{eq:SFP} on every $\Omega^Q_\alpha$. Substituting for the time derivatives shows that the boundary conditions are second-order, and this is why conditions are needed on every boundary and not only those with co-dimension 1. We call this set of equations the ``sticky'' equations because it is the Fokker-Planck equation for Sticky Brownian Motion \cite{ikeda81},
%(**ref friedlin+weber2003?), 
a stochastic process that has a probability atom on the boundary of its domain.

\section{Transition Rates for Sticky Brownian Motion}

\begin{figure*}
\centerline{\includegraphics[width=1\linewidth]{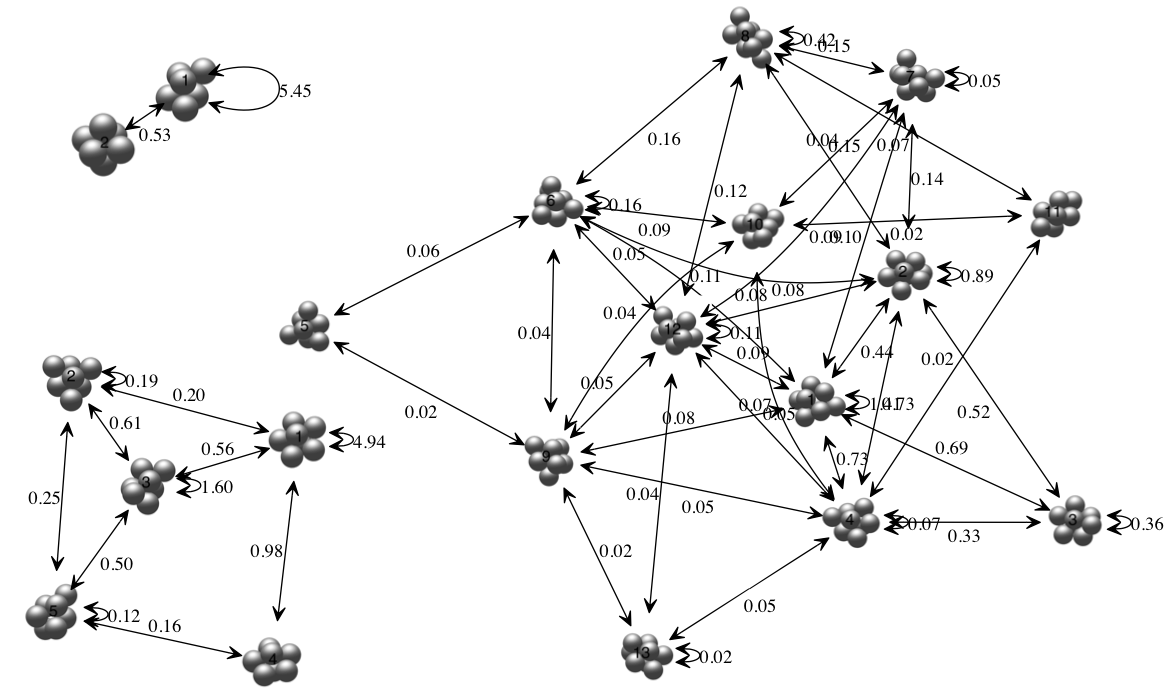}}
\caption{Rates (geometrical components) at leading-order, for $n=6,7,8$. Dimensional rates are $\kappa^{-1}D/d^2$ times the above. These rates indicate the total number of each type of transition one expects to see, per unit time.}
\label{fig:ratenetwork}
\end{figure*} 

\begin{figure}
\begin{minipage}{0.48\linewidth}
Theoretical counts\\

\begin{tabular}{ccc}
Mode & 1 & 2 \\
1 &  1570 $\pm$ 78   &  153 $\pm$ 24 \\
2 &   153 $\pm$ 24  & 0 \\
\end{tabular}
\end{minipage}\hfill
\begin{minipage}{0.48\linewidth}
Simulation counts\\

\begin{tabular}{ccc}
Mode & 1 & 2 \\
1 &  1256  & 124 \\
2 &  124  & 0 \\
\end{tabular}
\end{minipage}\\
\vspace{0.5cm}

\includegraphics[width=0.48\linewidth]{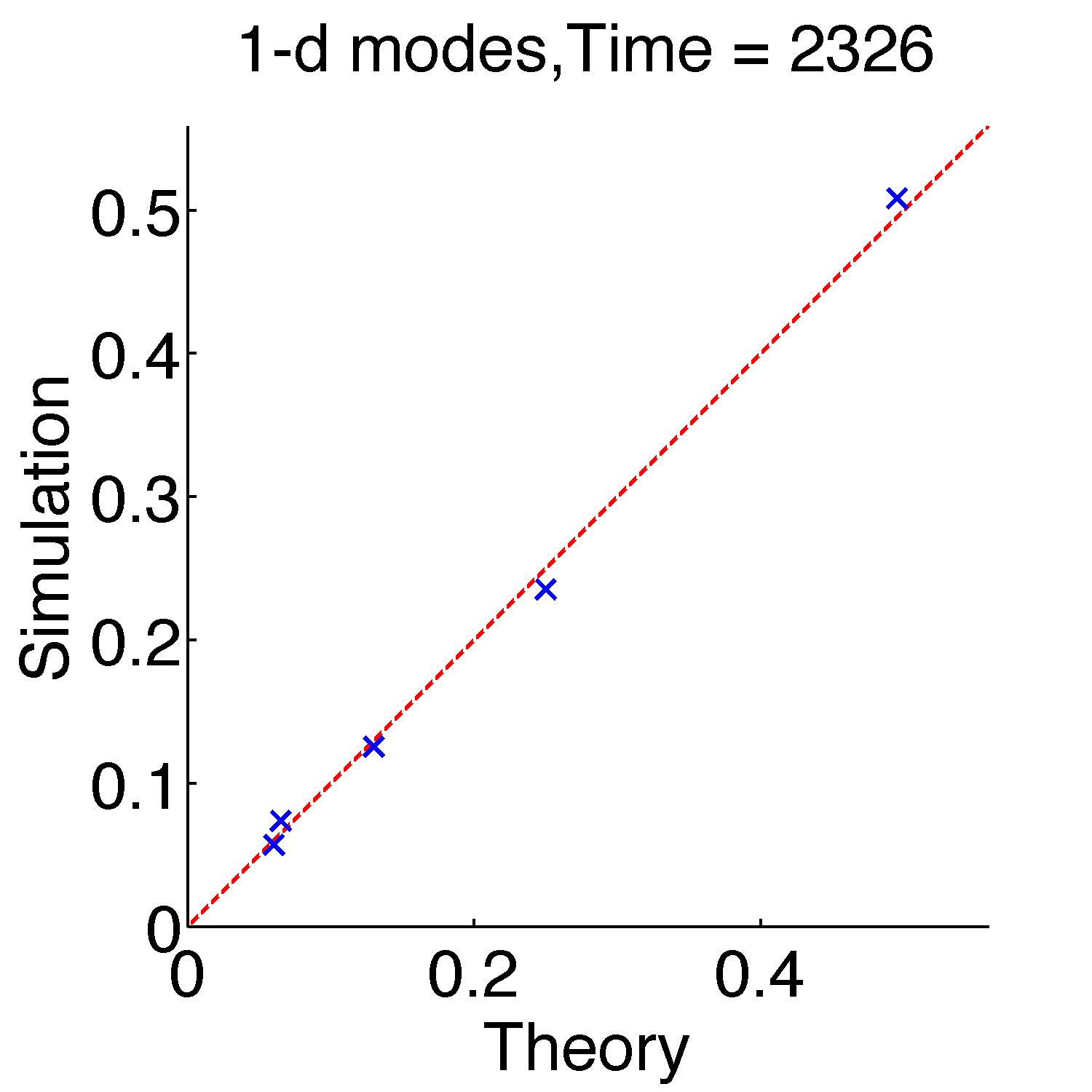}
\includegraphics[width=0.48\linewidth]{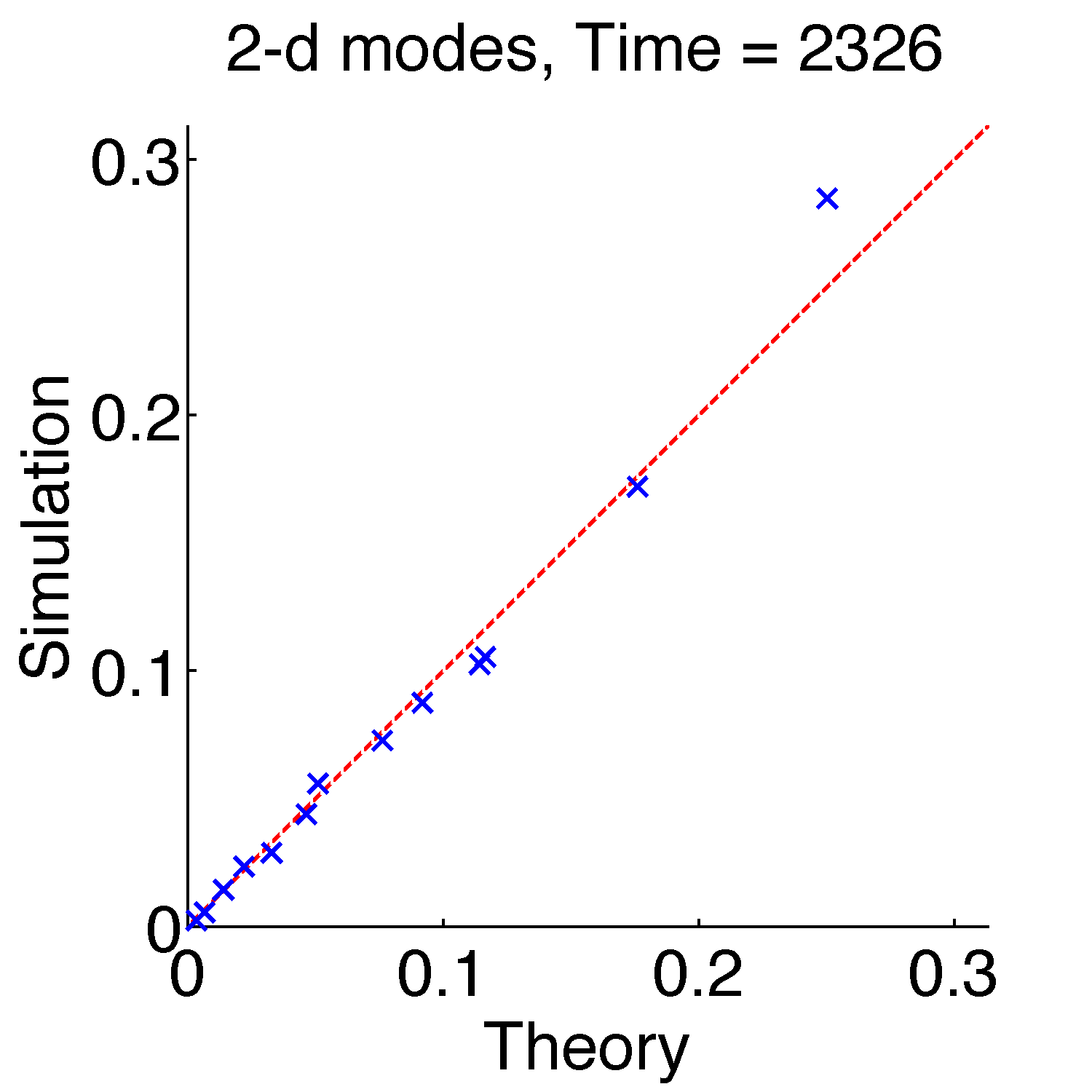}\\
\caption{Figures: Theoretical probabilities of each mode (relative to modes of the same dimension), versus probabilities computed from simulations using a potential with width $\approx 5\%$ of particle diameter,  for $n=6$. The red line is Theory=Simulation indicating a perfect match. Tables: number of transitions of each type, for simulation and theoretical prediction. The theoretical prediction indicates expected 95\% confidence intervals, computed using a normal approximation to a binomial. The sticky parameter was $\kappa = 16$ and the total running time was $2.3\times10^3$ units. 
}
\label{fig:sims}
\end{figure}

To transition between the rigid configurations in the geometrical limit, a cluster must break one (or more) bonds, then diffuse across the line segment (or face, etc) until it hits the other endpoint. The time it takes to do this depends on the length of the line and on how the entropy of the configuration varies along the line, and we can find this by solving an equation on the full line segment (face, etc). In our asymptotic limit there are no meaningful transition ``states'' -- rather, the entire line segment can be thought of as a transition state. Once the energetic barrier of breaking a bond has been overcome, transitions are dominated by diffusion. 

We now consider how to compute transition rates using the sticky equations \eqref{eq:SFP}, supposing we have a stochastic process $X(t)$ whose probability evolution is well-approximated by these.\footnote{
Note that we have not actually constructed a process that satisfies this in the strong sense. However, we began with a process satisfying \eqref{eq:dx} and showed the sticky equations describe its probability evolution  asymptotically so we use these to compute rates.}
Consider the transition rate between sets $A\subset\Omega$, $B\subset\Omega$, and 
for simplicity let us focus only on the case where these are both (disjoint) subsets of the 0-dimensional manifolds. 
For example, we may be interested in the transition rate between the octahedron and the polytetrahedron (introduced in Figure \ref{fig:schematic}), in which case $A$ would contain all points in the quotient space representing the octahedron, and $B$ all those representing the polytetrahedron. 
These rates can be computed using Transition Path Theory, which provides a mathematical framework for computing transition rates directly from the Fokker-Planck equations. We will simply state the facts that are relevant to our example and refer the reader to other resources for more details \cite{metzner06, eve2006,e2010}).

The \emph{committor function} $q(x)$ is the probability, starting from point $x$, of reaching set $B$ before set $A$. This solves the stationary backward Fokker-Planck equations with boundary conditions $q(A)=0$, $q(B)=1$, plus any other boundary conditions remaining from the equations. As shown in the Appendix, the forward sticky equations \eqref{eq:SFP} are self-adjoint (with respect to the invariant measure) so these are also the backward sticky equations. 

A \emph{reactive trajectory} is a segment of the path $X(t)$ that hits $B$ before $A$ going forward in time, and $A$ before $B$ going backward in time. The probability current of reactive trajectories is a vector field that, when integrated over a surface element, gives the net flux of reactive trajectories through it. Because our process is time-reversible, this current is \cite{eve2006}
\begin{equation}\label{eq:fluxreactive}
J(x) = \rho(x)\grad q(x) ,
\end{equation}
where $\rho(x)$  is the equilibrium probability measure. From \eqref{eq:Za} we find this is
\begin{equation}\label{eq:rho}
\rho(x) = Z^{-1}\sum_\alpha \kappa_\alpha(x)\delta_\alpha(x),
\end{equation}
where $\delta_\alpha(x)$ is the singular measure on $\Omega_\alpha$, i.e. it satisfies $\int_\Omega f(x)\delta_\alpha(x) = \int_{\Omega_\alpha} f(x)$.
Finally, the transition rate $k^{AB}$ is calculated by integrating this flux over a surface $S$ dividing the two states, giving
\begin{equation}
k^{AB} = \int_S J\cdot \hat{n} \; dS ,
\end{equation}
where $\hat{n}$ is the normal pointing from the side containing $A$ into the side containing $B$.

Computing this exactly requires solving the backward equations on a high-dimensional space, and integrating over a high-dimensional surface -- a computationally infeasible proposition. However, when the sticky parameter $\kappa$ is large, most of the probability is concentrated on the lowest-dimensional manifolds so we expect these to contribute the most to the rates. Therefore, let us expand the equations in powers of $\kappa^{-1}$. We suppose all variables have an asymptotic expansion as 
\begin{equation}
k^{AB} = k^{AB}_0 + \kappa^{-1}k^{AB}_1 + \ldots, 
\end{equation}
and similarly for $q$, $\mu$, $\rho$, $J$, etc. 
Expanding $\rho$ shows that to first order it is
%$\rho_0 = \mu_0 / Z_0$ which is 
a measure on points, to second order it is 
%$\rho_1 = \mu_1 / Z_0 - \mu_0Z_1/Z_0^2$ which is 
a measure on points and lines, etc. Measures on points will not contribute to the rate because the dividing surface $S$ can be chosen to avoid them, so $k_0^{AB} = 0$ and the leading-order part of the rate is $O(\kappa^{-1})$, computed from $\rho_1\grad q_0$. 

Expanding the backward sticky equations in powers of $\kappa^{-1}$ gives a set of equations for $q_0$:
\begin{equation}\label{eq:q0}
\divt{\alpha} h_\alpha I \gradt{\alpha} q_0 = 0 \qquad \text{on } \Omega_\alpha \quad(p >0),
\end{equation}
with boundary conditions $q_0(A) = 0$, $q_0(B) = 1$, and $\sum_{\beta\to\alpha} (h_{\beta}I\gradt{\beta}  q_0)\cdot \hat{n}^{\beta\alpha}=0$ on all other 0-dimensional manifolds. 

To solve these equations we can first find the solution on the 0- and 1-dimensional manifolds, then use this as a boundary condition for the solution on the 2-dimensional manifolds, which becomes in turn a boundary condition for the solution on 3-dimensional manifolds, etc. The leading-order rate  requires only  the solution for $p=0,1$. If we enumerate the lines connecting a point $a_k\in A$ to a point $b_k \in B$ and  use an arc-length parameterization for the $k$th line whose total length is $s_k$, then this given analytically as 
$q_0(s) = Q_{k}^{-1}\int_{s=0}^{s} \left(h_\alpha(s')I(s') \right)^{-1}ds', $
where 
\begin{equation}\label{eq:q0s}
Q_{k} = \int_{s=0}^{s_k} \left(h_\alpha(s') I(s') \right)^{-1}ds'
\end{equation}
 is the normalization factor. On all other lines $q_0(s) = 0$. 

Any dividing surface $S$ hits each line at a single point, so the leading-order rate (in dimensional units) is
\begin{equation}\label{eq:k1}
k_{AB}=\kappa^{-1} k_1^{AB} = \kappa^{-1}\frac{D}{d^2}Z_0^{-1}\sum_k Q_k^{-1},
\end{equation}
where the sum is over all connecting lines. This is asymptotically equivalent to the rate one would obtain simply by restricting the full committor function and invariant measure to the set of 0- and 1-dimensional manifolds.

\subsection{Transition Rates for hard spheres}

We used the formalism to compute the leading-order rates for $n=6,7,8$ hard spheres with diameter $d=1$. To do this we took the set of 1-dimensional solutions computed as part of the free energy landscape, and computed the factor $Q_k$ from \eqref{eq:k1} on each manifold. Summing over all of the 1-dimensional manifolds that connect a 0-dimensional manifold numbered $a$ to a 0-dimensional mode numbered $b$, gives the transition rate between $a$, $b$. We also include transitions between different ground states belonging to the same mode (e.g. $a\to a$), by multiplying the previous calculation by 2 since transitions can go in either direction along the line. 

Figure \ref{fig:ratenetwork} shows shows the network of 0-dimensional states and the reaction rates between each state. The numbers reported are the  dimensionless, purely geometrical  parts of the rates $Z_0^{-1}\sum_k Q_k^{-1}$, and should be multiplied by $\kappa^{-1}D/d^2$ to give the dimensional rate. These rates, when multiplied by the total time of a simulation or experiment,  give the average number of transitions one would expect to observe, so they are equal for both directions $a\to b$ and $b\to a$ since our system is time-reversible. 
 To obtain the rate relative to a particular state, i.e. the rate at which one leaves state $a$ to visit state $b$ next, given that the last state visited was state $a$, one should divide by the so-called reactive probability of $a$ \cite{schutte2011}. To leading order, this is equivalent to dividing by the equilibrium probability of $a$.

\paragraph{Simulations}

We have verified our results by performing Brownian dynamics simulations of \eqref{eq:dx} with a short-range Morse potential. The results agree very well with our calculations of both free energy and transition rates. 
Figure \ref{fig:sims} shows a comparison of the simulated probabilities versus theoretical probabilities of each mode for $n=6$ (see SM for $n=7,8$), for particles interacting with a Morse potential with range parameter $\rho=30$, a range of $\approx$ 5\% of the particle diameter. The agreement is nearly perfect. Fig. \ref{fig:sims} also compares the number of each type of transition we saw in the simulations, to that predicted from theory. The theory slightly overpredicts the total number of transitions -- this is what one should expect from the geometrical picture, as these leading-order rates neglect the time the system spends in the floppy manifolds, which would tend to slow it down. 

These results are encouraging partly because they are evidence that we have executed these calculations correctly, but  also because they suggest the asymptotic limit may apply for experimental systems, such as Meng et. al. \cite{meng2010} where the potential reportedly had roughly this width.

\subsection{Comparison with other numerical approaches to the free energy landscape}
We have compared our results with those obtained by a numerical study that directly searched an energy landscape of a short-ranged potential (a Morse potential, with range roughly 0.05 particle diameters) for local minima and transition states \cite{calvo2012}. 
The numerical method found fewer local minima than there are 0-dimensional modes, and fewer transition states than 1-dimensional modes,
 suggesting that the asymptotic theory has the roughest landscape.
 By computing adjacency matrices for the numerical states with a cutoff bond distance of 1 + 1$\times10^{-2}$, we verify that for $n=6,7,8$ each local minimum corresponds to a unique 0-dimensional mode, and each transition state lies on a unique 1-dimensional manifold.

We identify the point on the 1-dimensional manifold that is closest to each transition state.
The transition states are very close to the local maxima of the vibrational factor $-\log h(x)$ (see Appendix for plot), consistent with it being a saddle point of the potential energy. We believe the small discrepancy in location can be attributed to the finite width of the Morse potential used in the numerical procedure. 

The missing 0-dimensional modes occur for manifolds that are very close to each other in the quotient space metric (separation $\approx 0.08$), which is the case for modes $\{1,4\}$  ($n=7$) and modes $\{1,2,3,4\}$ ($n=8$). For these only one local minimum was found for the entire group.
 We hypothesize that because the separation is within the range of the potential, the 0-dimensional modes merge to form a single local minimum.
%Presumably this happens because the separation is within the range of the potential, so we don't expect the asymptotic theory to apply. 

The missing 1-dimensional modes often, but not always, correspond to self-self transitions  -- these transitions do not matter when particles are identical, however they will account for transitions between different states when the particles are not all the same (e.g. \cite{sahand2011}). For example, for $n=8$ the numerical procedure identifies 45 transition states, compared to our 75 one-dimensional manifolds. Of the missing manifolds, 16 are self-self transitions, 9 are nonself-nonself transitions within the group $\{1,2,3,4\}$, and 5 are nonself-nonself for endpoints not both in the group.

\section{Discussion / Conclusions}

We have developed a new framework for understanding energy landscapes when particles interact with a short-ranged potential. We show that in the limit as the range goes to zero and the depth goes to $-\infty$, the energy landscape becomes entirely governed by geometry, with a single parameter $\kappa$ encapsulating details about the potential and temperature. When $\kappa$ is large, only the lowest-dimensional geometrical manifolds contribute significantly to the landscape and this makes a computational approach tractable. 
To illustrate the limit, we have computed the set of low-dimensional manifolds for  $n\leq 8$ hard, spherical particles. 
This  solution to a nontrivial problem in statistical mechanics  can be used 
to compute equilibrium or non-equilibrium quantities for any potential whose range is short enough.

We were able to calculate this set of low-dimensional manifolds because we began with the set of rigid clusters, and made the conjecture that all floppy modes can be accessed from these by breaking bond constraints. Solving for the complete set of rigid clusters is a difficult problem in discrete geometry that has only been done for $n\leq11$ \cite{arkus2011,hoy2012}, but with current computational power and novel approaches \cite{hoy2012,wampler2005}  one can anticipate reaching larger $n$. Very large $n$ will eventually require making approximations to the geometry problem.
We speculate that as $n$ increases, structures with extra bonds, as well as ``singular'' structures whose Jacobians have extra zero eigenvalues, will come to dominate the landscape -- these have not yet been considered in our asymptotic framework but they are observed with high probability in experiments \cite{meng2010}.

 We have compared our results to those obtained by numerically searching the free energy landscape of a short-ranged Morse potential for local minima and transition states. Our method finds more minima and transition regions of the potential energy than the numerical search procedure, and this points to a potentially useful extension of our theory -- one can imagine starting with the limiting geometrical manifolds, and following these in some way as the range of the potential is increased, to obtain a low-dimensional approximation to the free energy landscape for  finite-width potentials, such as Lennard-Jones or Van der Waals clusters. This method would overcome a major issue with numerical searches which is that there is no way to ensure that all important parts of the landscape has been found -- we claim that our manifolds are the complete set of low-energy states so they will remain so under small enough perturbations. In addition, this would provide a way to deal with the increasing ruggedness of energy landscapes with short-ranged potentials, which are a challenge for numerical methods -- we \emph{start} with the most rugged landscape and would only need to smooth it. 

%Finally, we remark that the limit is in fact more general than the situation discussed here. For example, non-identical particles can be treated by allowing $\kappa$ to depend on the manifold, while deleting certain interactions altogether as in Hormoz et. al.\cite{sahand2011} simply changes the multiplicities $n_\alpha$ --  the manifolds do not have to be re-computed. One can also imagine incorporating long-range forces, non-spherical particles, potentials that vary in slow as well as fast coordinates, etc; and the limit applies to stochastic processes moving on a landscape even if they can't be mapped to particles. 

\begin{acknowledgments}
We thank David Wales for generously providing data, and Vinothan Manoharan and Eric Vanden-Eijnden for helpful discussions. This research was funded by the National Science Foundation
through the Harvard Materials Research Science and Engineering Center
(DMR-0820484),  the Division of Mathematical Sciences (DMS-0907985)
and the Kavli Institute for Bionano Science and Techology at Harvard University.
\end{acknowledgments}

-----------------------------------------------------------

%%%%%%%%%%%%%%%%%%%%%%%%%%
%%%    Bibliography    %%%
%%%%%%%%%%%%%%%%%%%%%%%%%%

%\bibliographystyle{pnas}%  
%\bibliographystyle{plainnat}  
%\bibliographystyle{aipsamp}
\bibliography{/Users/mirandaholmes-cerfon/Dropbox/ToFromHarvard/Bibliographies/ColloidBib.bib}
%\bibliography{/Users/miranda/Dropbox/ToFromHarvard/Bibliographies/ColloidBib.bib}

%%%%%%%%%%%%%%%%%%%%%%%%%%
%%%    Figures, Tables   %%%
%%%%%%%%%%%%%%%%%%%%%%%%%%

\clearpage

%%%%%%%%%%%%%%%%%%%%%%
%%%       Supplementary Material        %%%
%%%%%%%%%%%%%%%%%%%%%%

\appendix*

\section{Parameterization of a neighbourhood of $\Omega_\alpha$}

We provide a brief argument for the following statement made in the main text: given a regular point $x\in\Omega_\alpha$, there exists a differentiable parameterization of a neighbourhood $\mathcal{N}(x)\subset\R^{3n}$ of the form $y\times \{y_{\alpha_i}\}_{i=1}^m$, with $y\in\R^{3n-m}$, such that $\grad y\cdot \grad y_{\alpha_i} = 0$ on $\Omega_\alpha$. 

Recall that a \emph{regular} point $x\in\Omega_\alpha$ is a point such that the Jacobian of the transformation $x\to (y_{\alpha_1}(x), \ldots, y_{\alpha_m}(x))$ has rank $m$. 
If $x$ is regular then it has a neighbourhood $\mathcal{N}_\alpha(x)\subset\Omega_\alpha$ that is a differentiable manifold with co-dimension $m$ \cite{brucegiblin1984, lee2009}, so there exists a parameterization near $x$ by $y\in\mathbb{R}^{3n-m}$; let the associated mapping be $z:\R^{3n-m}\to\Omega_\alpha$. 

Given some point $x'\in\Omega^\eps_\alpha$ (not necessarily on $\Omega_\alpha$), we define a mapping  $x'\to (y_{\alpha_i},y)$ as $(y_{\alpha_i}(x'), y(x'))$, where $y(x')$ is found from the limit of the gradient flow map, i.e. $y(x') = z^{-1}(\lim_{t\to\infty}\phi(t))$ where $\phi$ solves $\dd{\phi}{t} = -\grad \sum_i U(y_{\alpha_i}(\phi))$, $\phi(0)=x'$. (We abbreviate $y_{\alpha_i}$ to mean the full list of constraint variables.) Results from \cite{falconer1983} (see also \cite{katzenberger1991}) show that this map exists  and is smooth enough in a neighbourhood $\subset\Omega$ of $x\in \Omega_\alpha$, provided $\Omega_\alpha$ is regular and $U(y)$ is sufficiently smooth. Since the  mapping $x'\to (y_{\alpha_i},y)$ has full rank at $x\in\Omega$, it does also in a neighbourhood $\mathcal{N}(x)\subset \Omega$ and so by the Inverse Function Theorem it is invertible. 
The orthogonality at $\Omega_\alpha$ follows because this is a level set of the constraints. 

Note that while this provides the required parameterization in a neighbourhood of $x\in\Omega_\alpha$, we have not shown that it extends to the set $\Omega_\alpha^\eps$ (see \eqref{eq:omegaeps}). This is not a problem for the asymptotic calculations, as these remain valid if $\Omega_\alpha^\eps$ is replaced with an atlas of local parameterizations $\mathcal{N}(x)$, patched together with a partition of unity -- the asymptotics only require the local behaviour near $y_{\alpha_i}=0$ and are not sensitive to the cutoffs at $r_c$.

\section{Quotient space metric and equations}

In this section we give more details about the quotient space and metric structure on it, and show how these arise naturally from our equations. 
Although these facts are well-known in Riemannian geometry \cite{lee2009,gallot} and well-used in chemistry and mechanics \cite{abrahammarsden,landau,sethna}, we have not found a  reference dealing succinctly with our particular context so we collect the relevant facts and demonstrations here. 

\paragraph{The manifold structure on the quotient space} Recall that we defined  the quotient space associated with a manifold $\Omega_\alpha$ to be $\Omega^Q_\alpha = \Omega_\alpha / G$, where the $G = SE(3)$ is the Special Euclidean group, i.e. the group of rotations and translations of a cluster. Then $\Omega^Q_\alpha$ is a smooth manifold if the Lie group $G$ acts \emph{properly} and \emph{freely} on $\Omega_\alpha$ (\cite{abrahammarsden}, Prop. 4.1.23 p.266, \cite{gallot}). To act properly is a compactness  condition and it can be checked that it is satisfied for $SE(3)$. To act freely means that the only element $g\in G$ such that $g\cdot x = x$ is the identity. This is the case provided there is no cluster in $\Omega_\alpha$ such that the spheres all lie on a line; for floppy manifolds with up to two bonds broken this is true when $n\geq 5$. 

The \emph{orbit} of a cluster $x$ is the set of points of the form $g\cdot x$ for $g\in G$, and is written as $[x]$. Each orbit is identified as an element in $\Omega^Q_\alpha$ by the canonical projection 
\begin{equation}\label{eq:proj}
\pi: \Omega_\alpha \to \Omega_\alpha / G := \{ [x]: x\in\Omega_\alpha\} .
\end{equation}
This projection shows how to map the tangent spaces to each other, via the pushforward map. Let $\mathcal{T}_\alpha(x)$, $\mathcal{T}_\alpha^Q([x])$ be the tangent spaces at $x\in\Omega_\alpha$, $[x]\in\Omega^Q_\alpha$ respectively. The tangent vectors map as follows: if $c(t)\in\Omega_\alpha$ is a curve such that $c(0) = x$, then the tangent vector $c'(0)\in\mathcal{T}_\alpha(x)$ maps to the tangent vector $\dd{}{t}\pi(c(t))|_{t=0} \in \mathcal{T}_\alpha^Q([x])$. 
Note that tangent vectors can also be identified as derivations, which we will write as $\partial_u$.

\paragraph{Metric on the quotient space}

The group  $G$ acts \emph{isometrically} on $\Omega_\alpha$, which means it respects the inner product $\langle \cdot,\cdot\rangle_{g_\alpha}$ on the manifold:  given $t_1, t_2 \in \mathcal{T}_\alpha(x)$, the inner product satisfies $\langle t_1, t_2 \rangle_{g_\alpha} = \langle g\cdot t_1, g \cdot t_2 \rangle_{g_\alpha} $ for all $g\in G$. 
Therefore one can construct a metric on the quotient manifold that is compatible with the projection
(making the projection a \emph{Riemannian submersion}), and this metric is unique \cite{gallot}.

To specify the metric we decompose $\mathcal{T}_\alpha(x)$ into the \emph{vertical subspace} $\mathcal{T}^v_\alpha(x)$, containing the directions tangent to the action of $G$, and the \emph{horizontal subspace} $\mathcal{T}^h_\alpha(x)$, its orthogonal complement. Therefore  $\mathcal{T}_\alpha(x) = \mathcal{T}^v_\alpha(x) \oplus \mathcal{T}^h_\alpha(x)$. 

Let $P_\alpha(x): \mathcal{T}_\alpha \to \mathcal{T}^h_\alpha$ the the orthogonal projection operator (we omit the argument $x$ for succinctness.) 
Let $[t]$ denote an element of $\mathcal{T}_\alpha^Q$ that has representative $t\in\mathcal{T}_\alpha$. 
The metric $\bar{g}_\alpha$ on the quotient space  $\Omega_\alpha^Q$ is computed from the metric $g_\alpha$ on $\Omega_\alpha$ as 
\begin{equation}
\langle [t_1], [t_2] \rangle_{g^Q_\alpha} =\langle \mathcal{P}_\alpha t_1, \mathcal{P}_\alpha t_2\rangle_{g_\alpha} .
\end{equation}

\paragraph{Fokker-Planck equations on the quotient space} Next we show how this differential structure arises naturally as a result of our manipulations to the Fokker-Planck equation. 

Consider a point $x_0\in\Omega_\alpha$,  %and a metric tensor $g_\alpha$, obtained by restricting the metric tensor in the ambient space to $\Omega_\alpha$. Let us choose the 
and let us parameterize a neighbourhood on $\Omega_\alpha$
in such a way that at $x_0$, the directions tangent to infinitesimal rotations and translations are orthogonal to the remaining variables. This can be done with the standard Euler angles. Let
\begin{eqnarray*}
R_x(\theta) &= 
\begin{pmatrix}
1 & 0 & 0 \\
0 & \cos \theta &  -\sin \theta \\ 
0 & \sin\theta & \cos\theta 
\end{pmatrix}\\
R_y(\theta) &= \begin{pmatrix}
\cos\theta & 0 & \sin\theta \\
 0 & 1 & 0 \\ 
 -\sin\theta & 0 & \cos \theta  \end{pmatrix} \\
 R_z(\theta) &= \begin{pmatrix}
 \cos\theta & -\sin\theta & 0 \\ 
 \sin\theta & \cos\theta & 0 \\ 
 0 & 0 & 1 \end{pmatrix} 
 \end{eqnarray*}
 be the matrices for rotation of a point about the $x,y,z$ axes respectively, 
with  block-diagonal versions appropriate to a cluster of $n$ particles $\mathbf{R}_x(\theta)$, $\mathbf{R}_y(\theta)$, $\mathbf{R}_z(\theta)$. These are obtained as, for example, 
 \[
 \mathbf{R}_x(\theta) = \begin{pmatrix}   
 R_x(\theta) & & \\
  & \ddots & \\
  & & R_x(\theta)
 \end{pmatrix}
 \]
 with $n$ copies along the diagonal and zeros everywhere else, and similarly for the other matrices. 
Let 
 \[
 \mathbf{T}(\mu_1,\mu_2,\mu_3) = (\mu_1,\mu_2,\mu_3,\mu_1,\ldots, \mu_1,\mu_2,\mu_3)^T
 \]
  be a vector representing translations. 
Let $\phi(y):\R^{p}\to\R^{3n}$ parameterize the remaining directions with $\phi(0) = x_0$, so that a neighbourhood of $x_0$ on $\Omega_\alpha$ can be parameterized with variables $(\theta_x,\theta_y,\theta_z,\mu_1,\mu_2,\mu_3,y)$ as
 \begin{equation}
 x = \mathbf{R}_z(\theta_z)\mathbf{R}_y(\theta_y)\mathbf{R}_x(\theta_x)\phi(y)
  + \mathbf{T}(\mu_1,\mu_2,\mu_3).
 \end{equation}
We can choose $\phi(y)$ so that it is orthogonal to infinitesimal rotations and translations at $x_0$: this means that we require $\pp{\phi}{y_k}\big\lvert_{y=0} \cdot \big(\pp{\mathbf{R}_x}{\theta_x}\big\lvert_{\theta_x=0}x_0\big) = 0$ , for each $y_k$ and each rotation matrix; the condition for translations is satisfied if the center of mass of $x_0$ is at the origin.  
%(Note that parameterizing rotations with a single chart will lead to problems at certain points, so the above formula should be taken somewhat lightly far away from $x_0$ as one should actually patch together two or more charts.)

The metric tensor on $\Omega_\alpha$ is $g_\alpha = J^TJ$ where $J$ is the Jacobian of the transformation $(\theta_x,\theta_y,\theta_z,\mu_1,\mu_2,\mu_3,y)\to x$. This has a simple block diagonal structure at $x_0$: 
\begin{equation}\label{eq:galph}
g_\alpha(x_0) = \begin{pmatrix}
\bar{g}_\alpha & 0 & 0 \\
0 & \mathbf{I} & 0 \\
 0 & 0 & I_3 
\end{pmatrix}
\end{equation}
where
\begin{equation}
 \quad \mathbf{I} = \begin{pmatrix} \sum_i y_i^2+z_i^2 & -\sum_i x_iy_i & -\sum_i x_iz_i \\ 
-\sum_i x_iy_i & \sum_i x_i^2+z_i^2 & -\sum_i -y_iz_i\\ 
-\sum_i x_iz_i & -\sum_i -y_iz_i & \sum_i x_i^2+y_i^2  \end{pmatrix} .
\end{equation}
Here $\bar{g}_\alpha$ is the $p\times p$ contribution from the $y$-variables,  $I_3$ is the $3\times 3$ identity matrix, and $\mathbf{I}$ is the moment of inertia tensor, which assumes the configuration is written as $x = (x_1,y_1,z_1,\ldots,x_n,y_n,z_n)$. We will write its determinant as $I^2(x_0) \equiv \det(\mathbf{I})$. 
  
Therefore  the Fokker-Planck equation on $\Omega_\alpha$  after integrating over the fast variables has the following form at $x_0$: 
\begin{widetext}
\begin{align}
\partial_t(\kappa^m h_\alpha p) &= \oneover{\sqrt{|g_\alpha|}}\partial_i\left( \sqrt{|g_\alpha|} \kappa^m h_\alpha g^{ij} \partial_j p\right) + \sum_{\beta\to\alpha}j_\beta\cdot\hat{n}^{\beta\alpha}  \nonumber\\
 &= \oneover{\sqrt{|\bar{g}_\alpha|I^2}}\partial_u\left( \sqrt{|\bar{g}_\alpha |I^2} \kappa^m h_\alpha \bar{g}^{uv} \partial_v p\right) 
 + \oneover{\sqrt{|\bar{g}_\alpha |I^2}}\partial_a \left( \sqrt{|\bar{g}_\alpha|I^2} \kappa^m h_\alpha g^{ab} \partial_b p\right) 
+ \sum_{\beta\to\alpha}j_\beta\cdot\hat{n}^{\beta\alpha} .
\label{eq:fp1}
\end{align}
\end{widetext}
Here $g_{ij}$, $g_{uv}$ are the elements of $g_\alpha$, $\bar{g}_\alpha$ respectively, $a,b$ index the rotational and translational variables, and we have substituted \eqref{eq:galph} in the second equation. We will not deal with the flux explicitly as this comes from  the same manipulations on the higher-dimensional manifolds. 

Integrating \eqref{eq:fp1} over orbits gets rid of the second term on the RHS, by Stokes' theorem, so we are left with
\begin{equation}\label{eq:fp2}
\partial_t(\kappa^m h_\alpha I \bar{p}) 
= \oneover{\sqrt{|\bar{g}_\alpha|}}\partial_u\left( \sqrt{|\bar{g}_\alpha |} \kappa^m h_\alpha I \bar{g}^{uv} \partial_v \bar{p}\right)  
+ \sum_{\beta\to\alpha}j_\beta\cdot\hat{n}^{\beta\alpha} . 
\end{equation}
where $\bar{p}$ is the integrated value of $p$ on an orbit (this is $Cp$ if $p$ is constant on the orbit, where the constant $C$ does not depend on $x_0$.) Because this does not depend on the location along the orbit of a point, we can identify it with a function on the quotient space as $\bar{p}(x,t) = \bar{p}^Q([x],t)$. For the same reason we can identify tangent vectors via the canonical projection as $\partial_u  \bar{p} = \partial_{[u]}\bar{p}^Q.$
The metric $\bar{g}_{uv}$ has the same elements as the quotient metric $\bar{g}_{[u][v]}$ because it only involves tangent vectors in the horizontal subspace, and it is independent of the representative $x_0$ that we chose for $[x_0]$ because $G$ acts isometrically.  

After identifying functions, tangent vectors, and the metric in \eqref{eq:fp2} with their projections in the quotient space, we obtain \eqref{eq:SFP} in the text. 

%This holds at every point $[x_0]$ although we do not construct a smooth parameterization.
%This is because the FP equation tells us the rate of change of $p$ as a function of its derivative along tangent directions. Once we know how it changes in one parameterization of the tangent space, we can change variables to write the equation in whatever parameterization we like. 

\paragraph{Representing the quotient space}

To parameterize the quotient manifold it is convenient to map it to a  space that has an explicit representation. For our numerical implementation we store the edge-lengths of all the particles (we call this ``bond-distance'' space.) This representation actually forms the quotient space with reflections as well, so it is only diffeomorphic to $\Omega^Q_\alpha$ if $\Omega_\alpha$ does not contain a cluster where all of the particles lie in a plane. 

An alternate parameterization would be to constrain one vertex to be at the origin, one vertex to lie on the $x$-axis, and one vertex to lie on the $xy$-plane. This would embed $\Omega_\alpha^Q$ in $\Omega_\alpha$. 

Given a parameterization of the quotient manifold in some space $B$ with canonical projection $\pi$, the tangent vectors and metric can be computed from the pushforward map \eqref{eq:proj} via numerical differentiation. That is, given a point $[x]\in B$ with representative $x\in \Omega_\alpha$ where this is two-dimensional, we compute the two unit tangent vectors $t_1, t_2 \in \mathcal{T}_\alpha(x)$ that are perpendicular to the infinitesimal rotations and translations. This is easy to do from the null space of the matrix $M$ defined in section \ref{sec:1dmanif}. We take small steps in each of these directions to obtain points $x_1 = x + \Delta s\: t_1$, $x_2 = x + \Delta s\: t_2$, project to $B$, and obtain first-order estimates of the quotient tangent directions as $[t_1] = (\pi(x_1)-x)/\Delta s$, $[t_2] = (\pi(x_2)-x)/\Delta s$. These have lengths $1$ and inner product $\langle t_1, t_2\rangle_{g_\alpha}$, which defines the metric on $B$. 

We use a first-order scheme to find the distance between two nearby points $[x_1], [x_2]$ in $B$. We find the separation vector $v = [x_1]-[x_2]$, project this onto the tangent space at $[x_1]$ and find the length of this projection. We repeat at the tangent space to $[x_2]$ and average the lengths. 

Note that parameterizing the quotient manifold as a subset of $\Omega_\alpha$ would imply slightly different numerical algorithms. For example, to find the distance between two nearby points, one would simply project the separation vector $v$ onto the horizontal tangent space at $x_1$, $x_2$ (these representatives can be chosen equal to $[x_1]$, $[x_2]$) -- this might be more efficient than the bond-distance space method.

\section{Adjoint equations}

In this section we compute the backward Fokker-Planck equation associated with \eqref{eq:SFP}. We could derive this using the same asymptotic procedure on the backward equation associated with \eqref{eq:dx}, however we prefer to demonstrate how to convert between the forward and backward sticky equations directly. We only outline the arguments here, leaving several steps to the reader. 

Recall that the backward equation describes the evolution of $u(x,t)$, the expected value of some function $g(x)$ that starts with a unit mass at $x$ and is subsequently stirred by the probability dynamics.
\footnote{If we were to construct a stochastic process $X_t$ with Fokker-Planck equation  \eqref{eq:SFP} and with initial condition $X_0 = x$, then we would be able to write $u(x,t) = \mathbb{E} g(X_t)$. Unfortunately we are only aware of such constructions for processes that have singular measures on manifolds of co-dimension 1 \cite{ikeda81}, and not processes that are sticky on manifolds of several different dimensions simultaneously, so we focus instead on the PDE interpretation.}
This is obtained from the transition probability measure $P_x(dy,t)$ as 
\begin{equation}\label{eq:u1}
u(x,t) = \int_\Omega P_x(dy,t)g(y) = \int_\Omega p(x,y,t)g(y)d\rho(y)
\end{equation}
where the transition probability has initial condition $P_x(y,0) = \sum_{\alpha}\delta_{\alpha}(y-x)$ and we have used the fact that it has a density $p(x,y,t)$ with respect to the equilibrium measure $d\rho(y)$ (see \eqref{eq:rho}). Note this implies $p(x,y,0)= \sum_\alpha\delta_\alpha(y-x)/\kappa_\alpha$. 

We suppose  that $P_x$ satisfies the Chapman-Kolmogorov equation
%\footnote{** Must show that $\int_z p(x,z,t-s)p(z,y,s)\mu(dz) = p(x,y,t)$, where $p$ satisfies FP equations in $y,t$. Not sure how to show this. Is it inherited from asymptotics? }, 
as this property holds for the original probability measure $p^\epsilon dx$ from which it derives: 
\begin{equation}\label{eq:CK}
\int_{z\in\Omega} P_x(dz,t-s)P_z(dy,s) = P_x(dy,t). 
\end{equation}
This allows us to write 
\begin{equation}\label{eq:uu}
u(x,t) = \int_\Omega P_x(dz,t-s)u(z,s) = \int_\Omega p_x(z,t-s)u(z,s)d\rho(z) .
\end{equation}
We can now obtain an evolution equation for $u$. Applying $\partial_t$ to \eqref{eq:uu} and using \eqref{eq:rho} gives
 \begin{align*}
\pp{u}{t} & = \int_{\Omega} p_tu \: d\rho(y) = \int_\Omega (\divt{}\gradt{}p)u\:d\rho(y) \\
 &= \int_{\Omega} p(\divt{}\gradt{} u) + \sum_i\int_{\Omega_i} u (\gradt{} p \cdot \hat{n}) \\
 & \qquad - p (\gradt{} u \cdot \hat{n}) + \kappa_i u \divt{}\gradt{}p .
\end{align*}
Here $\{\Omega_i\}$ is the set of manifolds of co-dimension 1 that form the boundary of the full space $\Omega$, $\kappa_i$ are the sticky factors along these manifolds, $\divt{}$, $\gradt{}$ denote differential operators on $\Omega$ and $\divt{i}$, $\gradt{i}$ will denote those on $\Omega_i$, $\hat{n}$ denotes a generic outward normal to the appropriate manifold, and integration is with respect to the volume element appropriate for each manifold. 
We now substitute for $\gradt{}p\cdot \hat{n}$ using the forward sticky equations to obtain
\begin{widetext}
\begin{align}
 \pp{u}{t} &= \int_\Omega p(\divt{}\gradt{} u) + \sum_i\int_{\Omega_i} u ( \divt{i} \kappa_i \gradt{i} p - \kappa_i\divt{}\gradt{}p) 
 %&\hspace{6cm} 
 - p\gradt{} u \cdot \hat{n} + \kappa_iu \divt{}\gradt{}p \nonumber\\
&= \int_\Omega p(\divt{}\gradt{} u) + \sum_i \int_{\Omega_i} p\left( \divt{i}\kappa_i \gradt{i} u - \gradt{} u \cdot\hat{n}  \right) %\nonumber\\
%& \hspace{4cm} 
+ \sum_j\int_{\Omega_j}\sum_{i\to j} \kappa_i (u\gradt{i} p \cdot \hat{n}^{(ij)} - p\gradt{i}u \cdot \hat{n}^{(ij)} ) \label{eq:ut}
\end{align}
\end{widetext}
Here $\{\Omega_j\}$ is the set of manifolds of co-dimension 2, forming the boundaries of the manifolds $\Omega_i$; the sum in the final term is over all manifolds $\Omega_i$ that have $\Omega_j$ as a boundary, and $\hat{n}^{(ij)}$ is the outward normal vector from $\Omega_i$ at $\Omega_j$. 

Evaluating \eqref{eq:ut} at $s=t$ gives the backward sticky equations
\begin{align}
\partial_t u &= \divt{}\gradt{}u  && \text{ in } \Omega \nonumber \\
\kappa_i\partial_t u &= \divt{i} \kappa_i \gradt{i} u - \grad u \cdot \hat{n} &&  \text{ in } \Omega_i \\
\partial_t u &= \text{(boundary terms)} && \text{ in } \Omega_j \nonumber
\end{align}
We stop at the first two terms; the interested reader can show that evaluating the boundary terms leads to the expected equations on the lower-dimensional manifolds. This set of equations is identical to the forward equations, so the system is self-adjoint.  

%The initial condition for the system above is $u(x,0) = g(x)\sum_i\kappa_i1_{\Omega_i}$, where $1_{\Omega_i}$ is the indicator function for each manifold. Therefore it is initially discontinuous. We expect the dynamics will smooth this out, as for the heat equation with discontinuous initial data, but some kind of weak theory should probably be done to verify this.

\section{Parameterizing the manifolds}

In this section we outline the method we used to parameterize the quotient manifolds  $\Omega^Q_\alpha$ describing clusters of hard spheres with up to 2 bonds broken. The method can be broken down into two separate sets of algorithms. The first algorithm generates points on the manifold, by taking linear steps along the tangent directions and projecting back down to the manifold. This is sufficient to calculate the 1-dimensional manifolds. The second algorithm links up the points with bars to form a simplex on which calculations can be performed, and is required for 2- and higher-dimensional manifolds. 

\subsection{1-dimensional manifolds}\label{sec:1dmanif}
Consider first the 1-dimensional manifolds. We take steps along the manifold as follows: given a point $x_0\in\Omega_\alpha$, a set of bond-distance constraints $\{y_k\}_{k=1}^m$, and a basis $\{t_i\}_{i=1}^6$ for the part of the tangent space at $x_0$ parallel to rotational and translational motions, we form a matrix $M = (\grad y_1, \ldots , \grad y_m, t_1, \ldots, t_6)^T$ and compute the null space of $M$. When $x_0$ is a regular point on the manifold,  this null space contains a single vector $v$ lying in the tangent space of $\Omega_\alpha(x_0)$, so we take a step in that direction as $x_1 = x_0+(\Delta s) v$. Because $v$ is orthogonal to translations and rotations, the length of our step in the quotient metric is $||x_1-x_0||_\alpha^Q = \Delta s + O(\Delta s^2)$. 

This step pushes us slightly off the manifold, so we project back down to it by finding a set of 
Lagrange multipliers $\lambda_k$ so that the projected point $x_1' = x_1 + \sum_k \lambda_k\grad y_k(x_1)$ lies on the manifold, i.e. we solve the nonlinear system of equations $y_k(x_1 + \sum_k\lambda_k\grad y_k(x_1)) = 0$ \cite{ciccottieve2008}. This is easily done using Newton's method as we are typically very close to the manifold. 

Beginning with a rigid cluster $x_0$ with an associated set of bond constraints, we break a bond by deleting one of the constraints, and perform the steps above until another bond is formed. This provides an ordered set of points in the quotient manifold, along with distances between them -- this is a ``line''.

\subsection{2-dimensional manifolds}
%(**see 0329 notes for detailed description)

Computing the 2-dimensional manifolds is more involved; here are the steps we followed. 

\paragraph{Compute the boundaries} 
First, we compute the boundaries of the manifold. To compute a 2-dimensional manifold $\Omega^Q_\alpha$ with constraint list $y_1(x), \ldots, y_n(x)$, we start with a corner point $x_0\in \Omega^Q_\alpha$ with two extra constraints $y_{i_1}(x), y_{i_2}(x)$. Deleting one of these, say $y_{i_1}$, we walk along the 1-dimensional manifold (as in the previous section) until another bond is formed, corresponding say to constraint $y_{i_3}$. We add this to our constraint list, delete the next extraneous constraint $y_{i_2}$, and repeat. We continue in this way, deleting one extraneous constraint at each corner, until we reach the original corner $x_0$. This gives us the boundary of $\Omega^Q_\alpha$, including corners. 

\paragraph{Generate points in the interior} 
Second, we generate a collection of points in the interior. Beginning with every point on the boundary, we generate ÒlinesÓ in the interior by holding all constraints fixed except the one that takes us off the boundary, and walk in this direction until we exit the manifold. We throw away points that are too close in some metric to existing points to avoid generating too many points.

\paragraph{Triangulate the points}
 Third, we triangulate the points. Many typical algorithms will not work here because our surface is not embedded in $\R^3$,  so we adopt an algorithm proposed by by \cite{floater2001} (see also \cite{floater1997}) that works as follows. 
\begin{enumerate}
\item Map the boundary to a fixed convex polygon in $\R^2$. 
\item  Map the interior points to the interior of that region in $\R^2$,  by letting each interior point be a convex combination of its neighbouring points. More specifically: let $x_1, \ldots x_n$ be the set of interior points, and let $N_i$ be a neighbourhood of an interior point $x_i$. Given a set of strictly positive weights $\lambda_{ij}$ such that $\sum_{x_j\in N_i} \lambda_{ij} =1$, find parameter points $u_1, \ldots u_n \in \R^2$ that solve the linear system of equations 
\[
u_i = \sum_{x_j\in N_i} \lambda_{ij}u_j, \qquad i=1,\ldots, n. 
\]
Each $u_i$ is contained in the convex hull of its neighbours, so it will be in the interior of the region defined in step (1) \cite{floater2001}. 
\item Triangulate the parameter points in $\R^2$ (we use a Delauney triangulation.) This lifts back to a triangulation of the manifold. 
\end{enumerate}

%Choosing the boundary, weights and neighbourhood is somewhat of an art. 

To implement this, we choose the boundary region so that the corners lie on a circle with a fixed radius, and the line segments joining them lie on arcs of circles whose lengths are approximately the same as the lines they are parameterizing. Points along these arcs are placed so the inter-point distance in the plane is proportional to the distance in the quotient metric between the points. The quality of the triangulation will depend on the choice of boundary region, and we find better qualities as the angles at the corners more closely represent the angles on the manifold.

Choosing the neighbourhood $N_i$ is a balance between sampling many points to get a smoother parameterization, and choosing fewer points so the manifold is roughly linear in the neighbourhood and does not contain any folds or other external branches of the manifold. We choose the neighbourhood to be the $k$ nearest points along the manifold, where a range of roughly $8 \leq k \leq 15$ works well for the step sizes we use, but $k$ will increase as step size decreases.

There are many ways to choose the weights; the most straightforward is for them to be the same, but almost as straightforward is for them to be inversely proportional to distance in some metric. For rapid but still good quality results we use the metric in bond-distance space -- this takes the list of pairwise bond distances and computes the Euclidean distance between the vectors.

\paragraph{Improve the triangulation}
 Finally, we improve the quality of the triangulation by letting the triangle sides be springs and evolving the points on the manifold with the spring forces, re-triangulating when necessary. 
Springs are chosen to be slightly longer than the average distance between points so the points want to spread out and fill the whole space. When a point hits a boundary it is absorbed, and is subsequently constrained to move along the boundary. This algorithm was introduced by \cite{persson2004} for a triangulation of the Euclidean plane, and we adapt it by replacing the length in the plane with the distance metric in the quotient manifold. In practice, we typically use bond-space distance instead as this is faster to compute and still gives a good quality triangulation. 

\paragraph{Integration on the manifolds} To compute quantities integrated over the manifolds, such as for the partition functions $z^{geom}_\alpha$, we used  finite elements on the simplex with standard piecewise linear elements. The sides of the triangles must be  calculated in the quotient space metric.

\subsection{Remarks}

\paragraph{Topology} The calculations above require that the 2-dimensional manifolds be topologically equivalent to a disc. 
%Our parameterization algorithm will detect when this is not the case, because we move the points around to fill the whole domain and identify the boundary they hit when they leave.\footnote{There is one slight caveat, which is that we represent points in the computer in bond-distance space. This is equivalent to forming the quotient space with reflections as well as rotations. We have checked by hand (via Geomags) that none of our floppy manifolds contain reflections, except one with $n=5$ particles that we treat separately. A future version of the algorithm will avoid this problem by representing points in Euclidean space by fixing a triangle of points.} 
Because we have obtained smooth parameterizations, we are confident that this is true for all the floppy manifolds under consideration. Alternatively, one could show this from a collection of points using Betti numbers \cite{silva2004}, for example. 

This points to an interesting question in discrete geometry -- under what conditions is the topology of floppy manifolds a polygon? One has to rule out surfaces of higher genus and non-orientable surfaces, among other things. We expect this can be shown for low-dimensional manifolds and small $n$, but larger $p$ or $n$ may be more complicated.

\paragraph{Numerical parameters and convergence} For the calculations reported in the text we used a step size of $\Delta s = 0.01$ to calculate the one-dimensional manifolds, and $\Delta s = 0.05$ to generate points on the boundary and interior of the 2-dimensional manifolds. When generating points we threw away points that were closer than $0.5\Delta s$ in bond-distance space to already-generated points. We typically ran the triangulation step 3 times before re-triangulating, using a step size of $\Delta t = 0.1$ and an internal pressure parameter of $1.2$ (see \cite{persson2004}), although a small selection of manifolds had to be re-triangulated after each spring step. We ran the triangulation until the area of the manifold, computed in bond-distance space, changed less than $\Delta s / 20$ after 3 consecutive re-triangulations. We checked for convergence in two ways: by calculating the  manifolds using a coarser resolution ($\Delta s = 0.1$ for the 2-dimensional manifolds, $\Delta s = 0.05$ for the 1-dimensional manifolds), and by running the triangulation algorithm for longer, until the minimum triangle quality ($q = 2r_{in}/r_{out}$, the ratio between (twice) the radius of the largest inscribed circle and the smallest circumscribed circle, see e.g. \cite{field2000}) was greater than 0.2. Both tests allowed us to conclude that our calculated ratios $Z_2/Z_1$, $Z_1,Z_0$ are correct to $\pm 0.05$, $\pm 0.01$ respectively  (although we have not reported this many decimal points for the latter).

\section{Simulations}

We have performed Brownian dynamics simulations of interacting particles to test our asymptotic calculations. 
We solve equation \eqref{eq:dx} with $D=\beta=1$ using a forward Euler timestep. For the potential we started with a Morse potential with maximum depth $E$ at $r=1$ and range parameter $\rho$; this takes the form $U(r) = Ee^{-\rho(r-1)}\left( e^{-\rho(r-1)}-2\right)$. The hard-core part for $r<1$ was modelled with a parabolic potential of the form $U(r) = \half m^2U''(1)(r-1)^2 - E$ for some number $m$; we choose $m=2$. This constrains the time step to be $\Delta t \ll (m^2U''(1))^{-1}$ and we typically choose a factor of 6--8 less. The sticky factor, accounting for the parabolic part, is $\kappa = \frac{m+1}{m}\frac{e^{E}} {\sqrt{2E\rho^2}}\sqrt{\frac{\pi}{2}}$. 

The whole potential was truncated at $r_c = 1+ 4/\rho$, by adding a linear term to keep the force continuous at $r_c$, as $U_{trunc}(r) = U(r) - ( U(r_c)+U'(r_c)(r-r_c))$ for $r<r_c$, and $U_{trunc}(r)=0$ otherwise. This modifies the sticky parameter to 
\begin{equation}\label{eq:kaptrunc}
\kappa = \frac{m+1}{m}\frac{\exp\left(E - U(r_c)-U'(r_c)(1-r_c)\right)} {\sqrt{2E\rho^2}}\sqrt{\frac{\pi}{2}}.
\end{equation}

We began with an initial condition drawn from the equilibrium distribution of rigid modes, and ran several copies of the simulation for a very long time. At time increments of $1\times10^{-2}$ we checked to see which floppy mode the cluster occupied  by forming its adjacency matrix, computing the number of bonds, and if this number was $\geq3n-6-2$, finding the adjacency matrix in our list of floppy modes to which it was topologically equivalent. 
This allows us to compute the occupation probabilities of each floppy mode.

To form the adjacency matrix, we said particles were bonded when they were at a distance of less than $1 + 2/\rho$; this is the range beyond which the force is negligible. (Choosing a bond distance of $1+1/\rho$ gave results that were inconsistent with the asymptotics.) We used the Matlab function \texttt{graphisomorphism.m} to determine topological equivalencies. 

We also kept track of transitions between rigid clusters, by recording the times and mode numbers at which the system first hits a rigid state that occupies a separate part of configuration space than the previous rigid state. This is easily done by checking whether or not the current adjacency matrix of a rigid mode is identical to the adjacency matrix of the previous rigid mode. 

\begin{figure}
\includegraphics[width=0.3\linewidth]{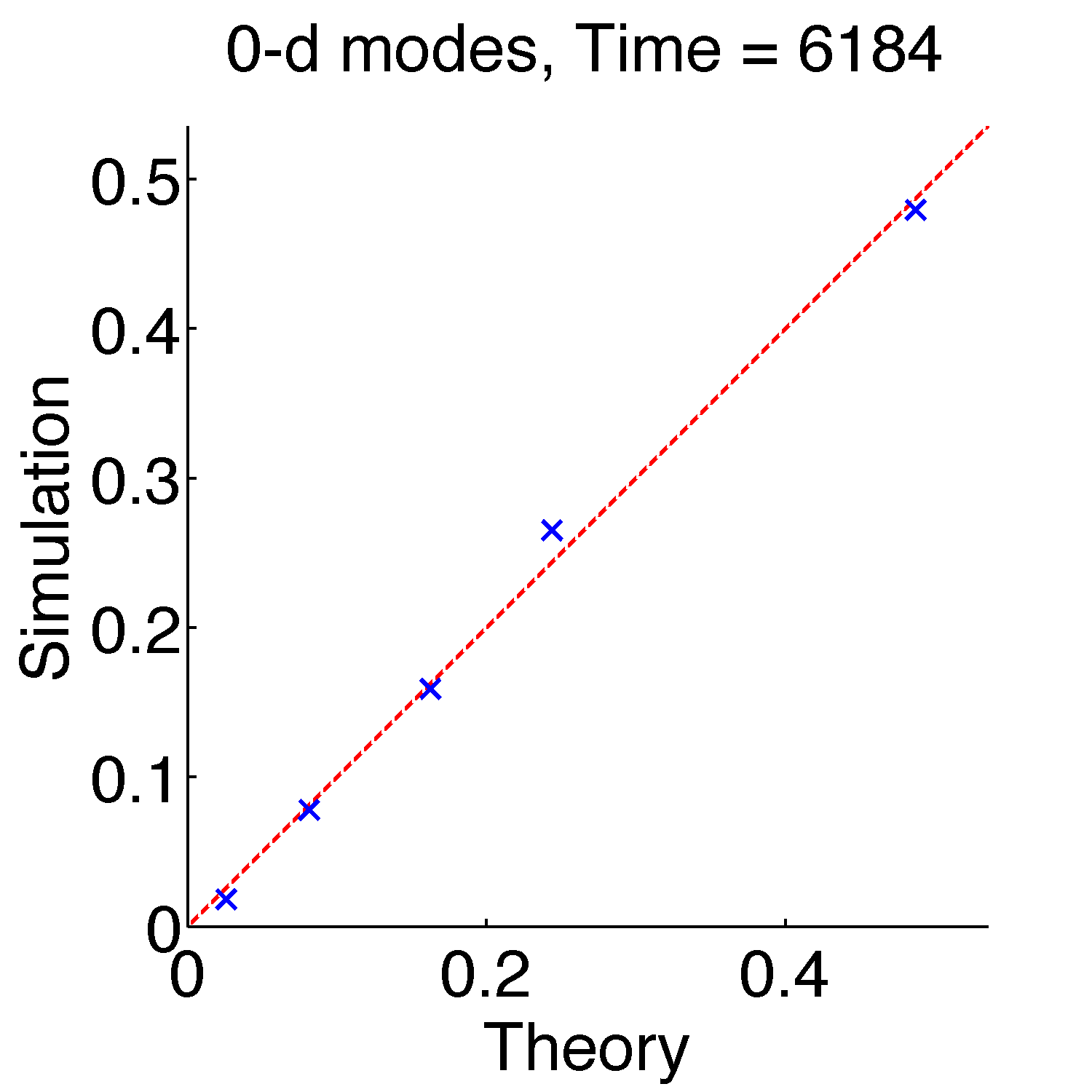}
\includegraphics[width=0.3\linewidth]{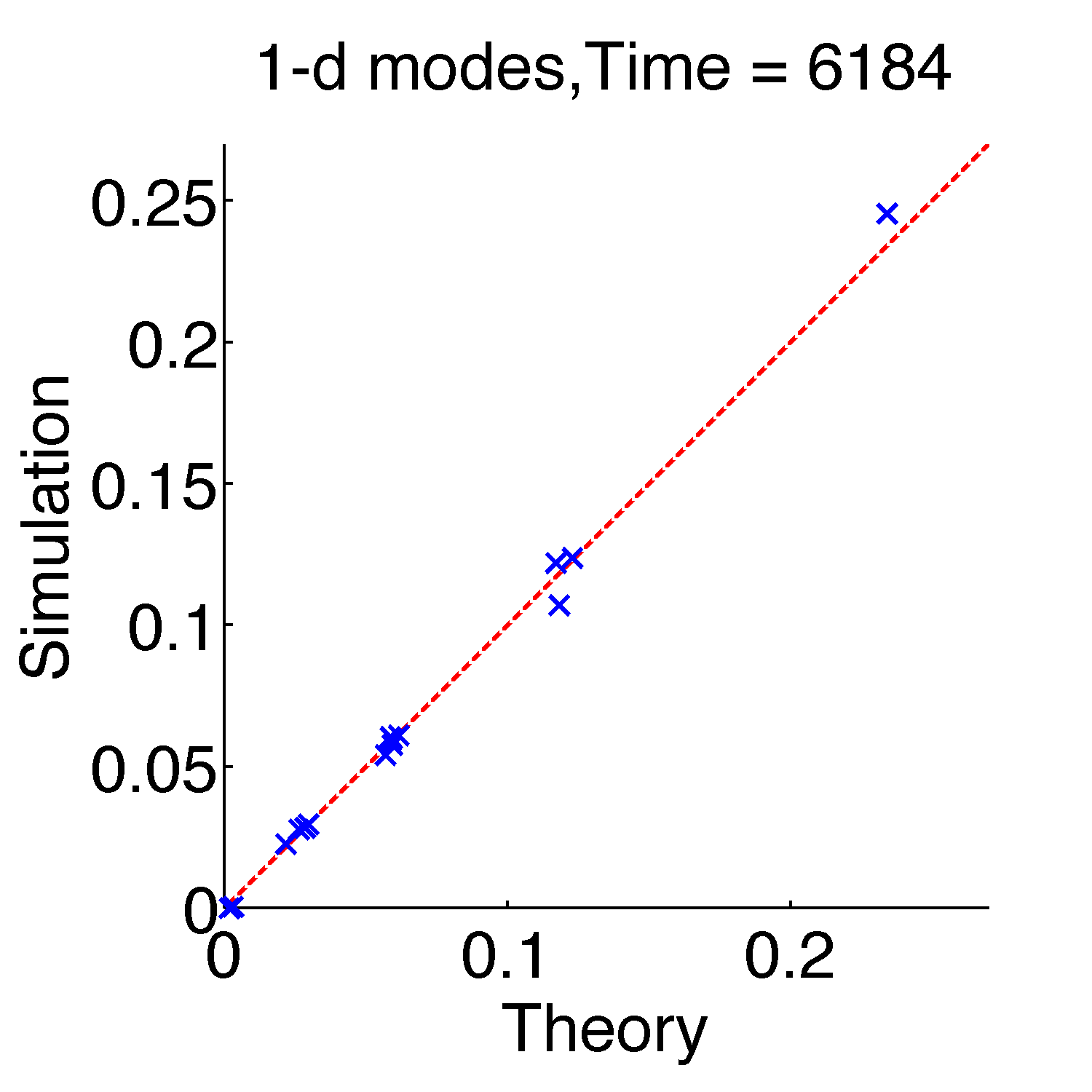}
\includegraphics[width=0.3\linewidth]{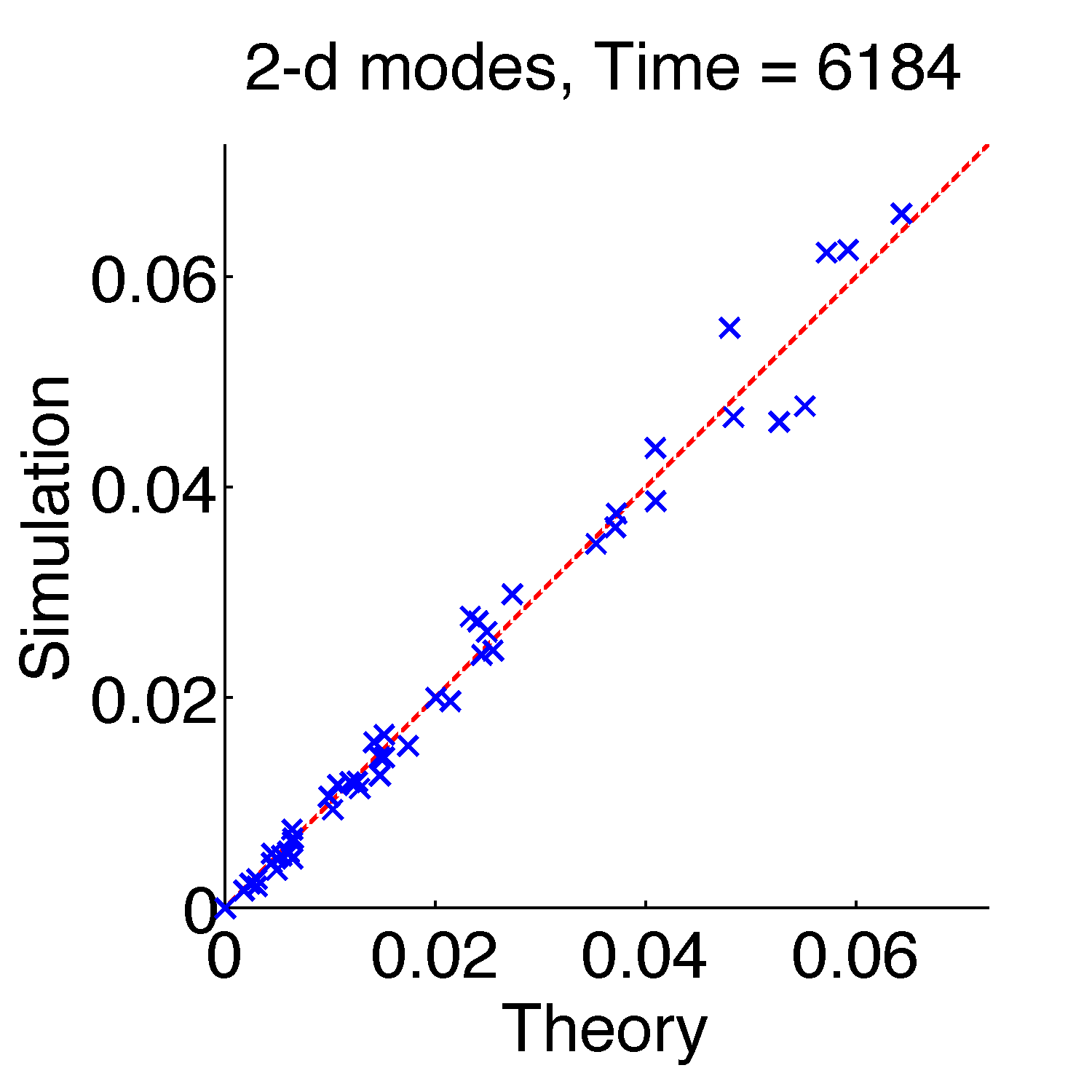}\\
\includegraphics[width=0.3\linewidth]{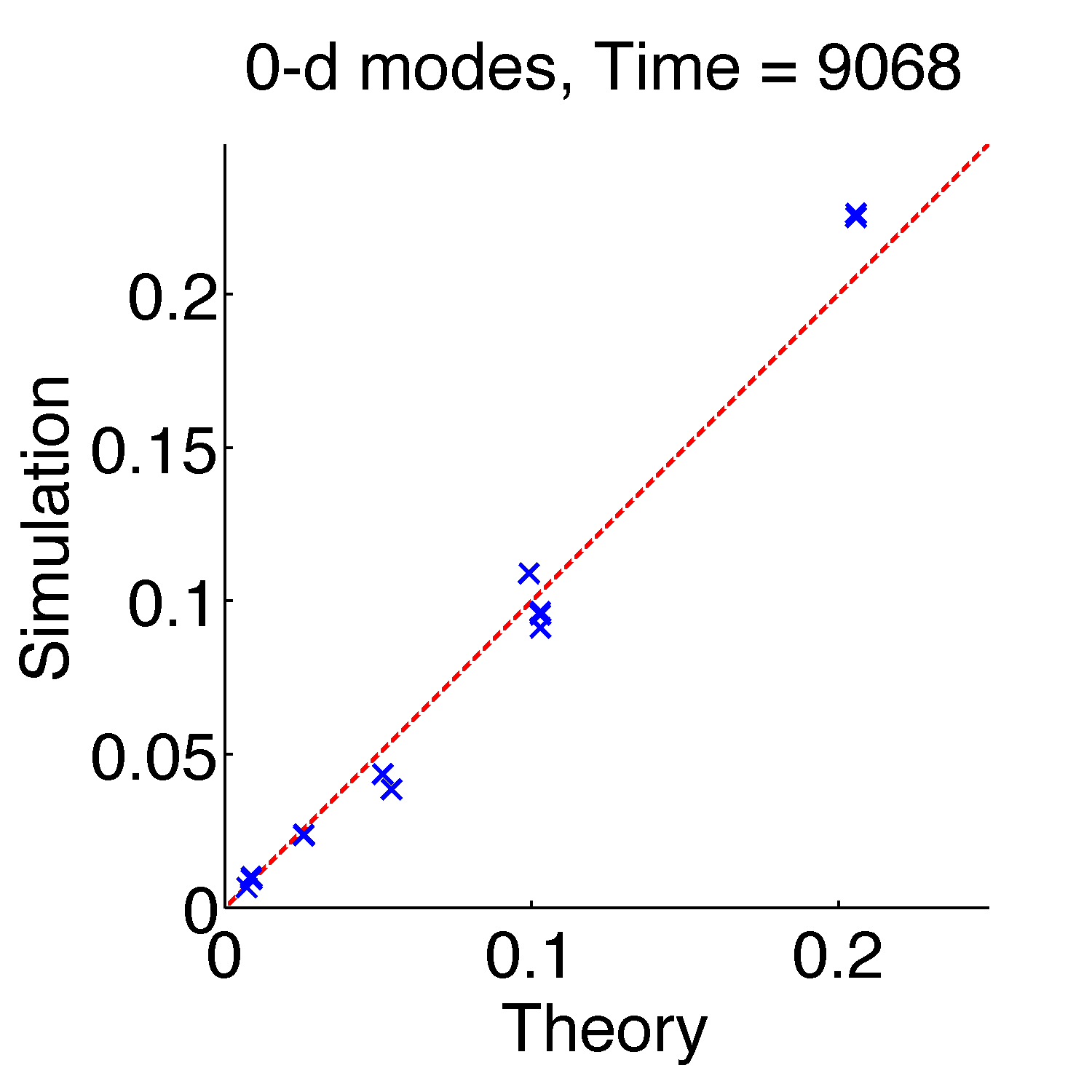}
\includegraphics[width=0.3\linewidth]{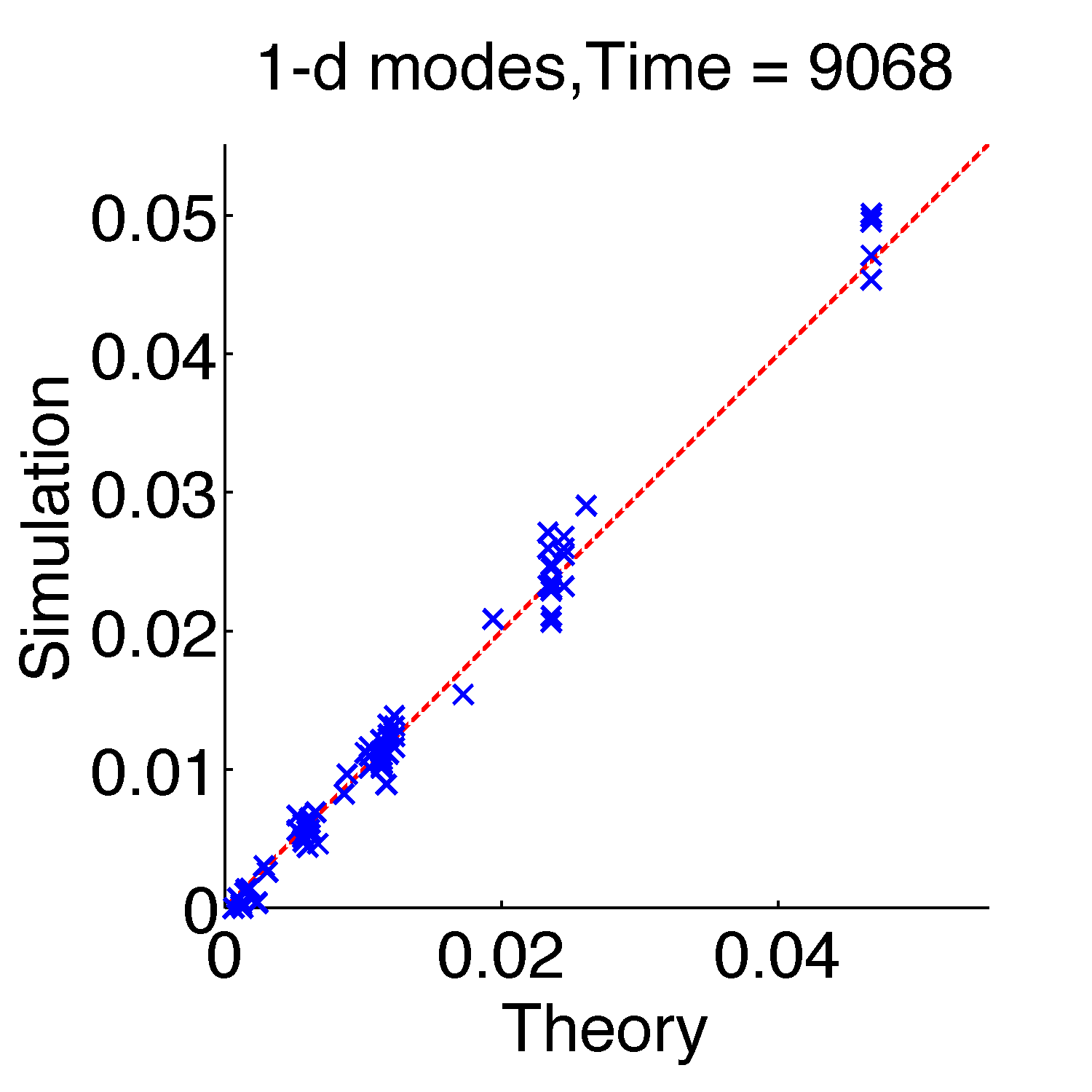}
\includegraphics[width=0.3\linewidth]{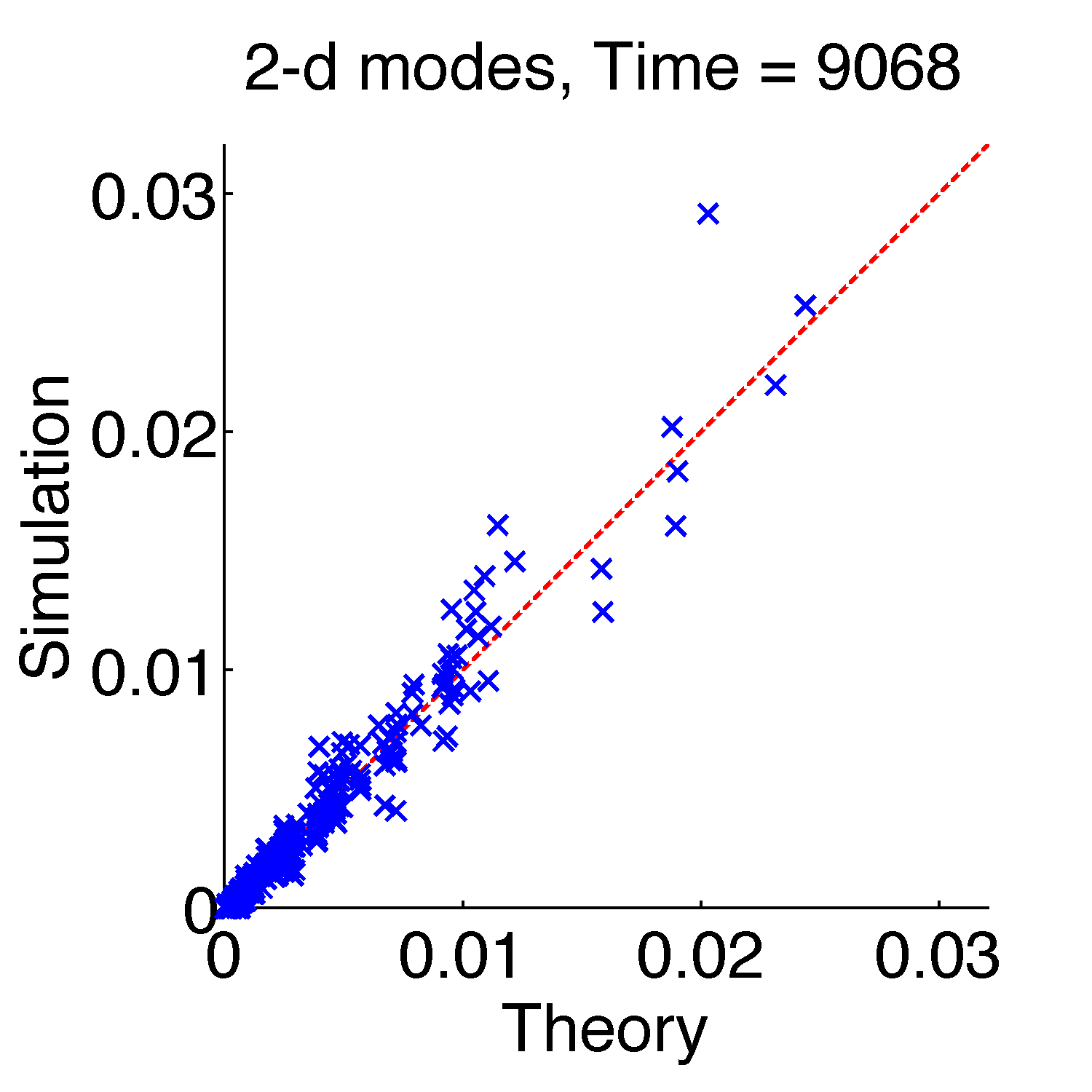}
\caption{
Simulated versus theoretical probabilities of floppy modes, for $n=7$ (top), $n=8$ (bottom), for a Morse potential with range parameter $\rho=30$. The running time of each simulation varied and is given in the title. 
 }\label{fig:sims78}
\end{figure}

Figure \ref{fig:sims78} shows the simulated probabilities versus theoretically computed probabilities of each floppy mode for $n=7,8$, for simulations using a Morse potential with the same parameters as in the text (Figure \ref{fig:sims}): $E=8.5$, $\rho=30$, $dt=2\times10^{-6}$. This implies the sticky parameter is $\kappa = 16$. Again, there is excellent agreement. 

\begin{figure}
\includegraphics[width=0.3\linewidth]{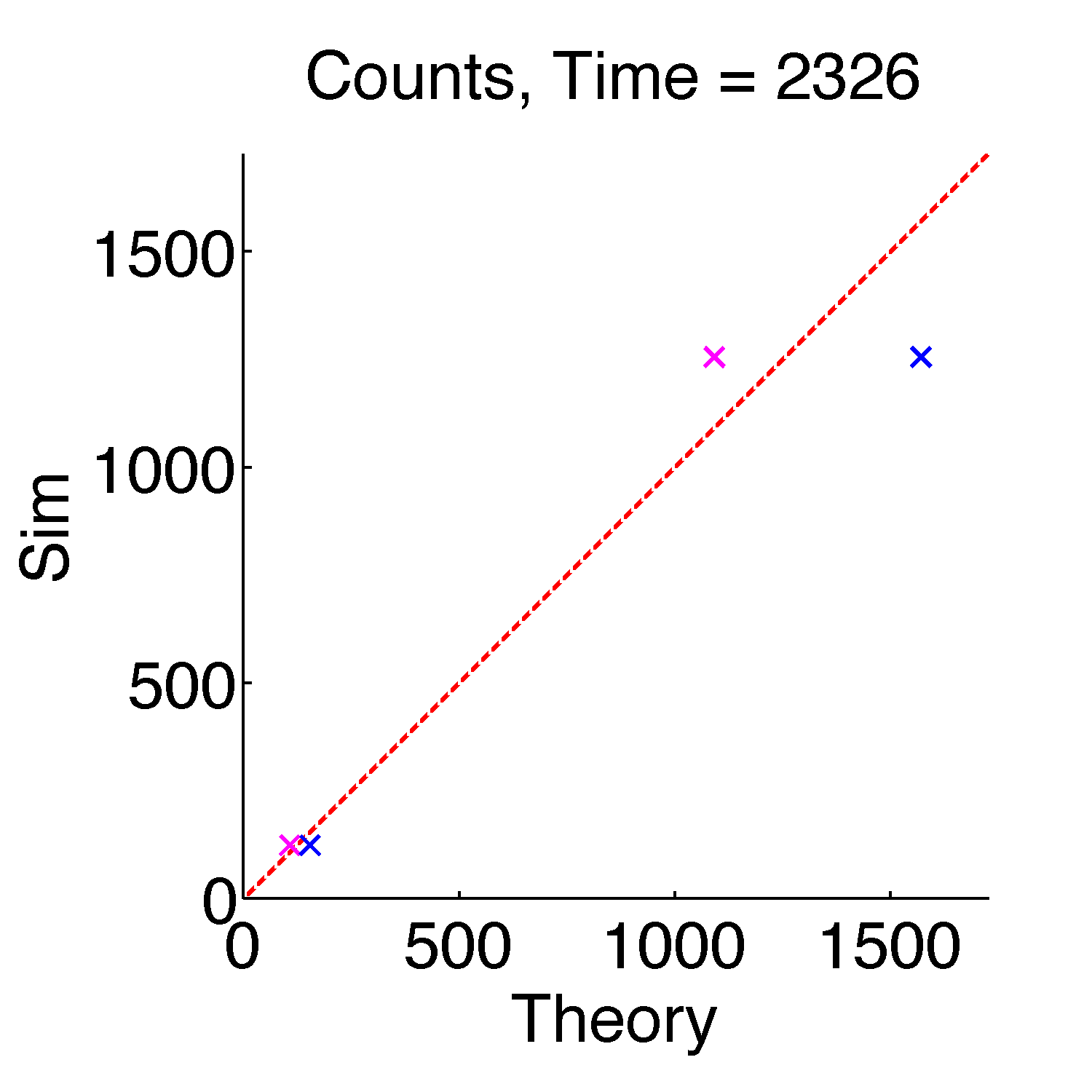}
\includegraphics[width=0.3\linewidth]{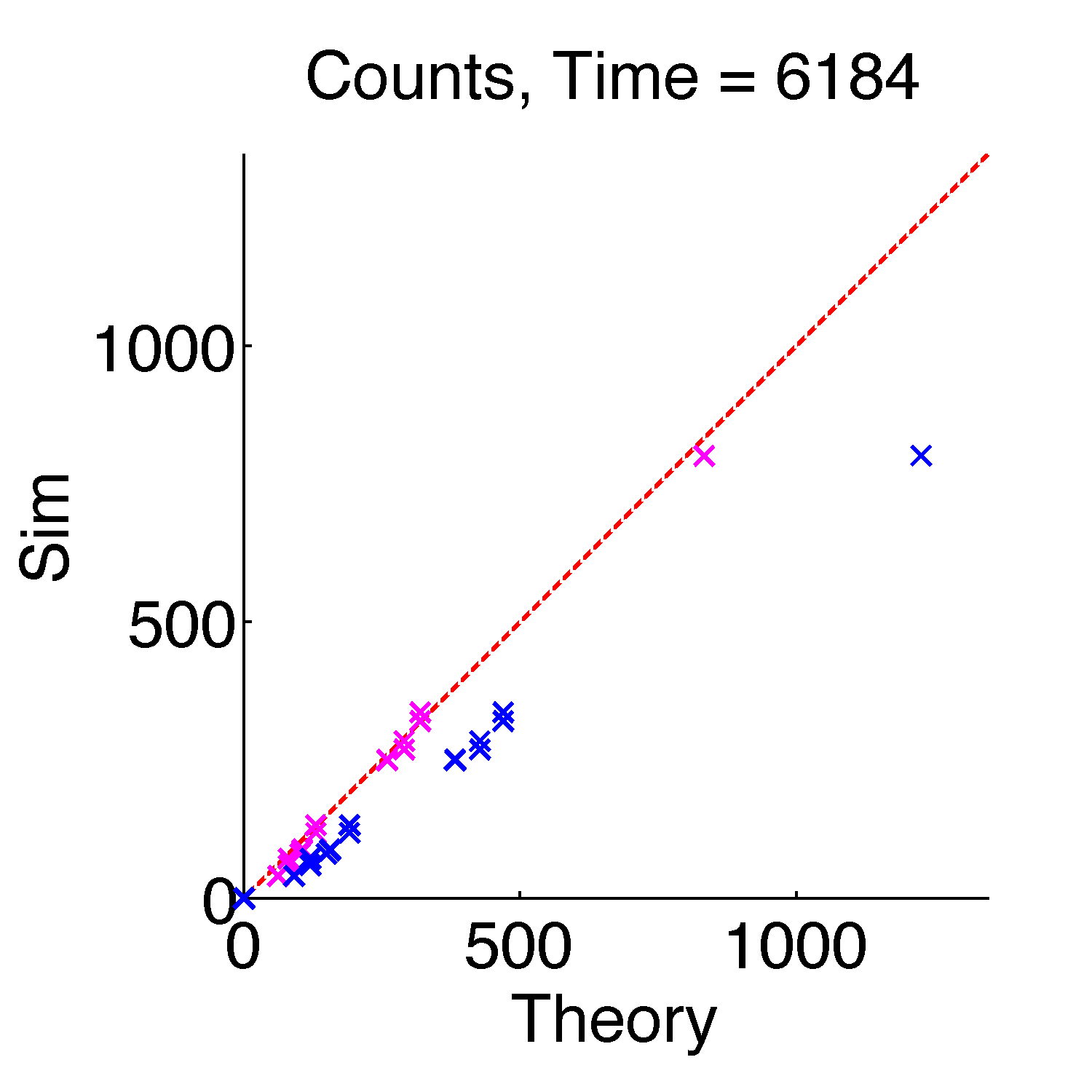}
\includegraphics[width=0.3\linewidth]{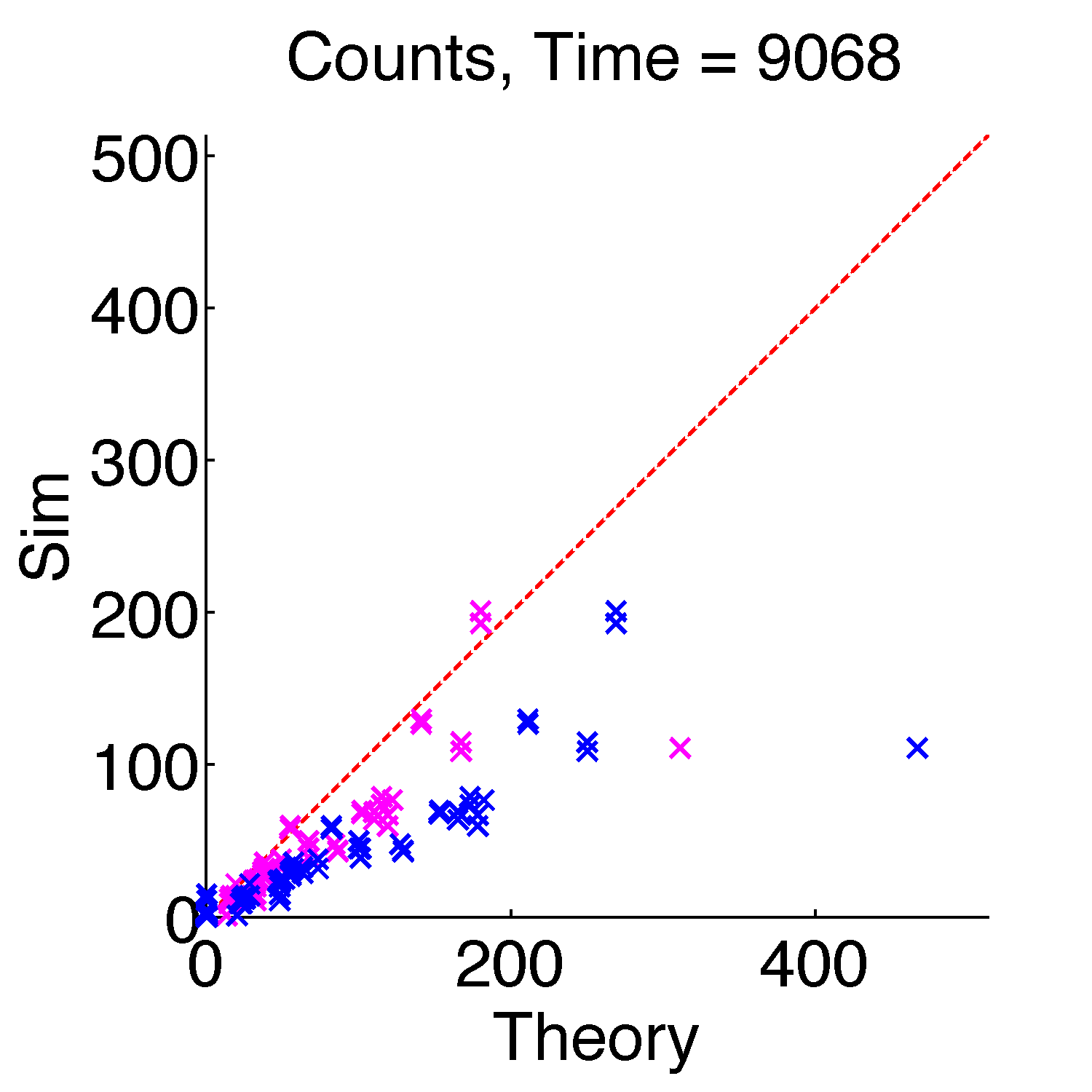}
\caption{Elements of the count matrix for $n=6,7,8$ (left, middle, right). 
}\label{fig:counts}
\end{figure}

Figure \ref{fig:counts} plots the elements of the transition count matrix for $n=6,7,8$ from simulations, versus two theoretical calculations. The blue markers come from the leading-order asymptotic approximation, computed from \eqref{eq:k1}. The magenta markers come simply restricting the dynamics to the network of points and lines, which gives geometric rates of $(Z_0+\kappa^{-1}Z_1)^{-1}\sum_k Q_k^{-1}$ -- these are the blue rates multiplied by $1 / ( 1+\kappa^{-1} Z_1/Z_0)$, so are asymptotically equivalent, but uniformly smaller.These rates do a better job of predicting the simulated rates for this value of the range parameter. 

To calculate rates for $n=7,8$ we have grouped the 0-dimensional modes that are very close together into one mode (these are modes $\{1,4\}$ for $n=7$, modes $\{ 1,2,3,4\}$ for $n=8$). The rate from a mode in this group to a mode not in the group is the sum of the rates out of each mode in a group, and the rates within a group are ignored. We were able to make a distinction between modes in a group only by using a range parameter of $\rho=150$ (not shown). 

Table \ref{tbl:ratios} shows the ratios $Z_{p+1}/Z_p$ extracted from the simulations. These are uniformly smaller than the theoretically computed values, but approach the theory as the range of the potential decreases. This is because the simulations use an excess bond length that is finite rather than infinitesimal, so some clusters are counted as being on $p$-dimensional manifolds, when in the limit they would be on $p+1$-dimensional manifolds. One could potentially correct for this if one had no knowledge of the potential, but wanted to use these measurements to estimate the sticky parameter.

\begin{table}
\begin{minipage}{0.48\linewidth}
$Z_1 / Z_0$\\
\begin{tabular}{cccc}
  & A & B & C \\
  6 & 5.5 & 5.9 & \\
  7 & 5.8 & 6.0 & 7.2 \\
  8 & 5.2 & 6.4 & 6.7\\
\end{tabular}
\end{minipage}
\begin{minipage}{0.48\linewidth}
$Z_2 / Z_0$\\
\begin{tabular}{cccc}
  & A & B & C \\
  6 & 3.4 & 4.1 & \\
  7 & 3.5 & 3.8 & 4.2 \\
  8 & 3.6 & 3.6 & 4.7\\
\end{tabular}
\end{minipage}
\caption{Ratios extracted from simulations with different range and sticky parameters. These were computed as (time in $p+1$-dimensional modes) / (time in $p$-dimensional modes) / $\kappa$, where $\kappa$ is computed using \eqref{eq:kaptrunc}. Parameters were (A) $dt = 2e-6$, $E = 8.5$, $\rho = 30$ ($\kappa = 16$), (B) $dt = 8e-7$, $E = 10$, $\rho = 50$ ($\kappa = 31$), (C)  $dt = 2.5e-7$, $E = 10.6$, $\rho = 150$ ($\kappa = 22$)}\label{tbl:ratios}
\end{table}

\begin{table}
\begin{tabular}{cccccrl}
Mode & $\bar{h}$ & $\bar{I}$ & S & $n_\alpha$ & $z_\alpha$ & corners\\
\hline
\multicolumn{4}{l}{0-dimensional modes} \\

   1 &      0.061 &       3.16 &  &        180 &      34.64  \\
   2 &      0.034 &       2.83 &  &         15 &       1.44  \\
\hline
\multicolumn{4}{l}{1-dimensional modes} \\

   3 &      0.066 &       3.30 &       0.85 &        180 &      33.30 &    1,  1  \\
   4 &      0.063 &       3.29 &       0.89 &         90 &      16.65 &    1,  1  \\
   5 &      0.057 &       3.29 &       0.95 &        360 &      64.03 &    1,  1  \\
   6 &      0.069 &       3.49 &       1.47 &        360 &     126.89 &    1,  1  \\
   7 &      0.045 &       3.04 &       0.63 &        180 &      15.40 &    1,  2  \\
\hline
\multicolumn{4}{l}{2-dimensional modes} \\

  16 &      0.075 &       3.37 &       0.35 &        180 &      15.99 &    1,   1,  1  \\
  17 &      0.075 &       3.76 &       2.00 &        360 &     202.45 &    1,   1,   1,   1,  1  \\
  18 &      0.083 &       4.01 &       2.17 &        180 &     130.10 &    1,   1,   1,  1  \\
  19 &      0.064 &       3.53 &       1.07 &        360 &      87.44 &    1,   1,   1,  1  \\
  20 &      0.057 &       3.17 &       0.23 &        180 &       7.52 &    1,   1,  2  \\
  21 &      0.073 &       3.79 &       2.84 &        360 &     284.23 &    1,   1,   1,   1,   1,  1  \\
  22 &      0.055 &       3.17 &       0.24 &         90 &       3.76 &    1,   1,  2  \\
  23 &      0.064 &       3.56 &       1.55 &         72 &      25.33 &    1,   1,   1,   1,  1  \\
  24 &      0.067 &       3.48 &       0.64 &        360 &      53.23 &    1,   1,  1  \\
  25 &      0.063 &       3.59 &       1.58 &        360 &     129.53 &    1,   1,   1,   2,  1  \\
  26 &      0.054 &       3.27 &       0.59 &        360 &      37.56 &    1,   1,   2,  1  \\
  27 &      0.081 &       4.07 &       2.67 &        120 &     105.52 &    1,   1,  1  \\
  28 &      0.072 &       3.77 &       2.39 &         90 &      57.97 &    1,   1,   1,  1  \\
\hline\\
\end{tabular}
\caption{Free energy data for each mode, $n=6$.}\label{tbl:data}
\end{table}

\begin{figure*}
\begin{center}
\includegraphics[width=0.2\linewidth]{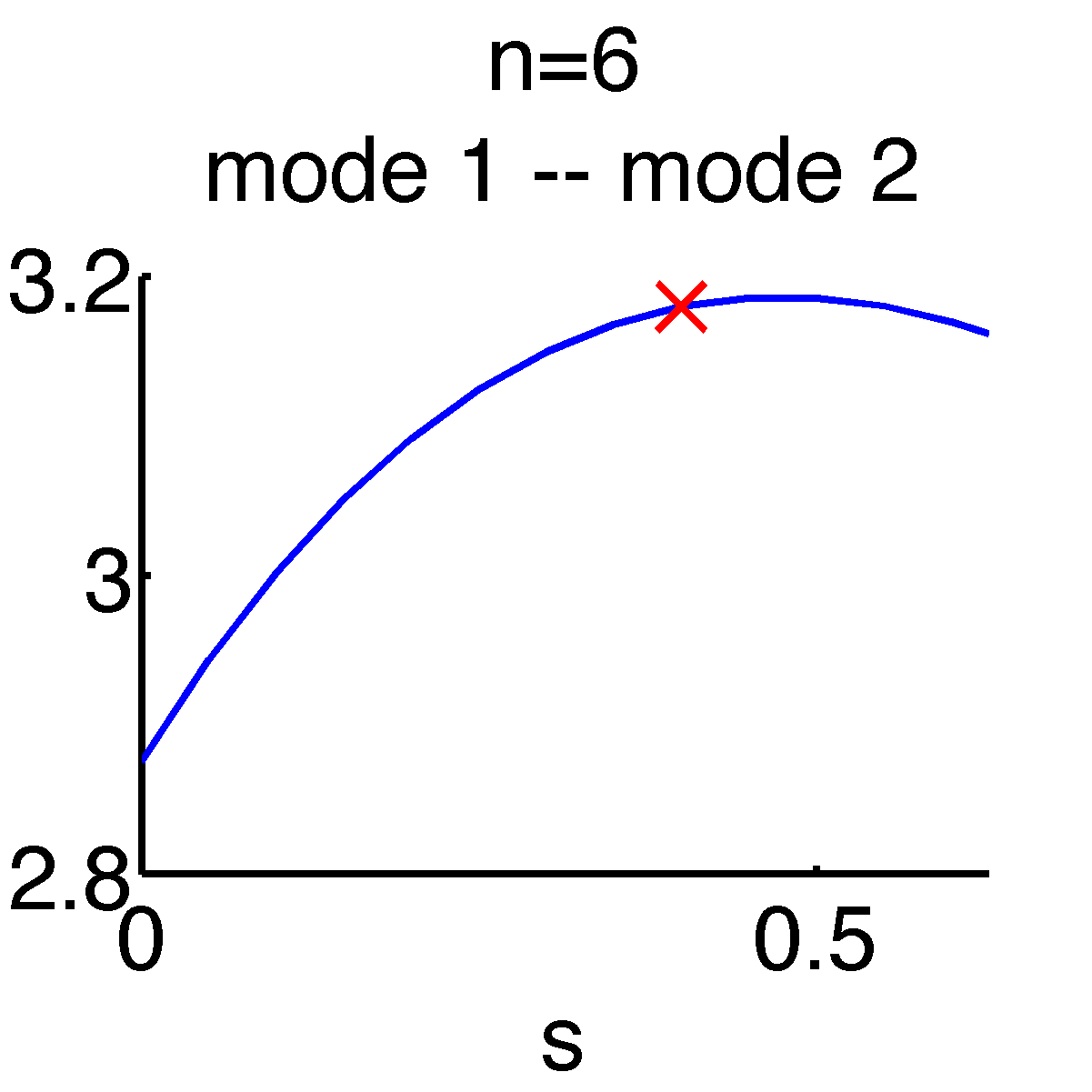}
\includegraphics[width=0.2\linewidth]{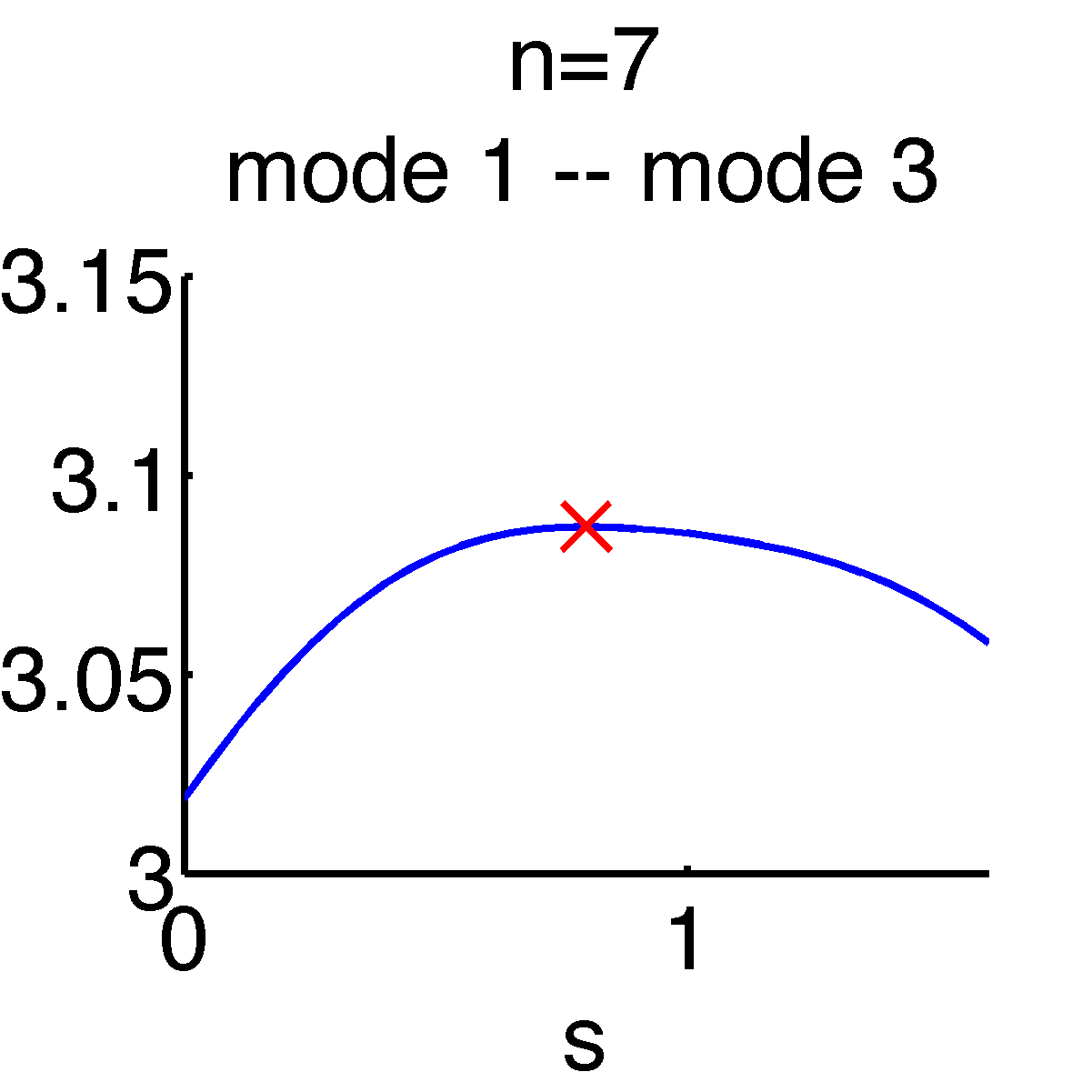}
\includegraphics[width=0.2\linewidth]{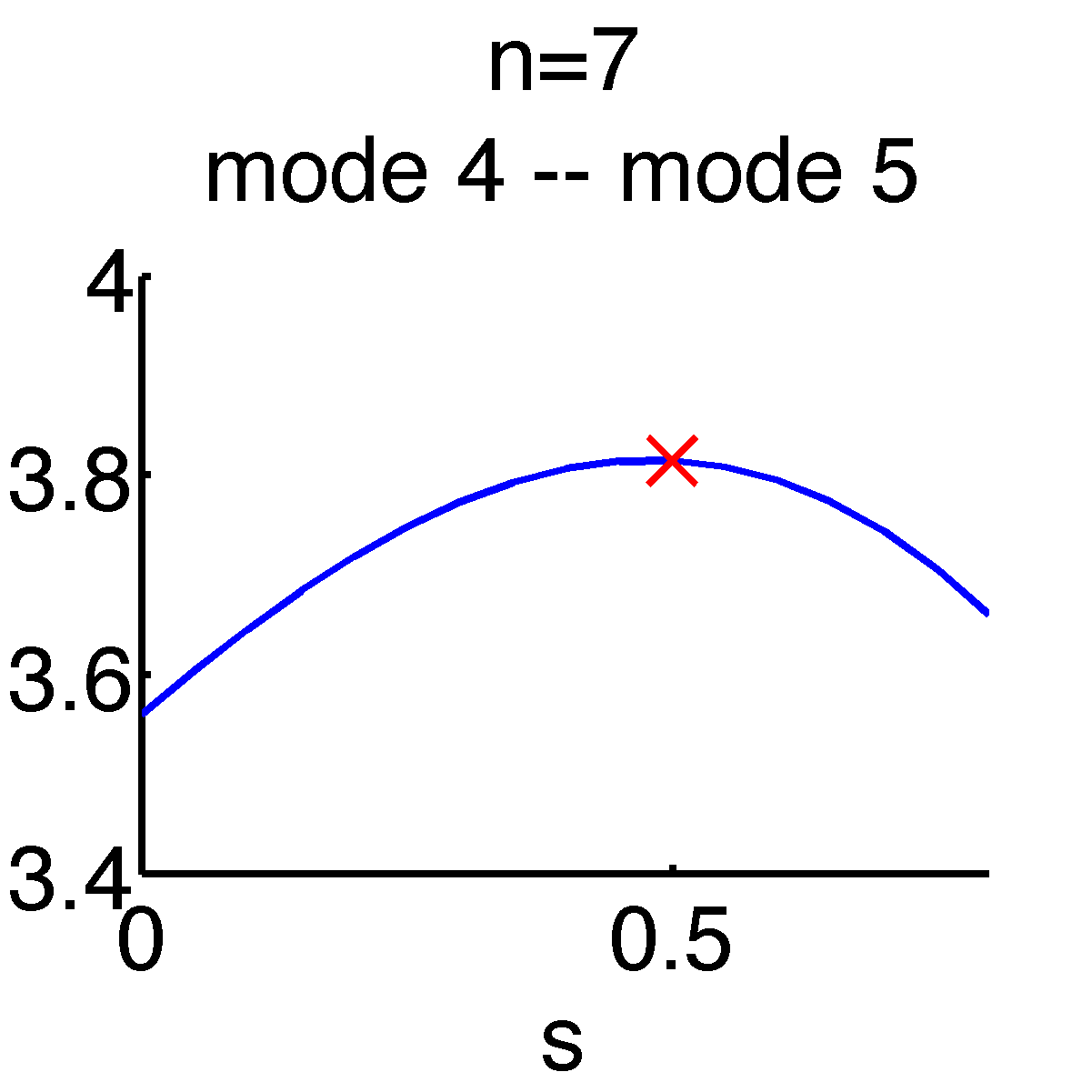}
\includegraphics[width=0.2\linewidth]{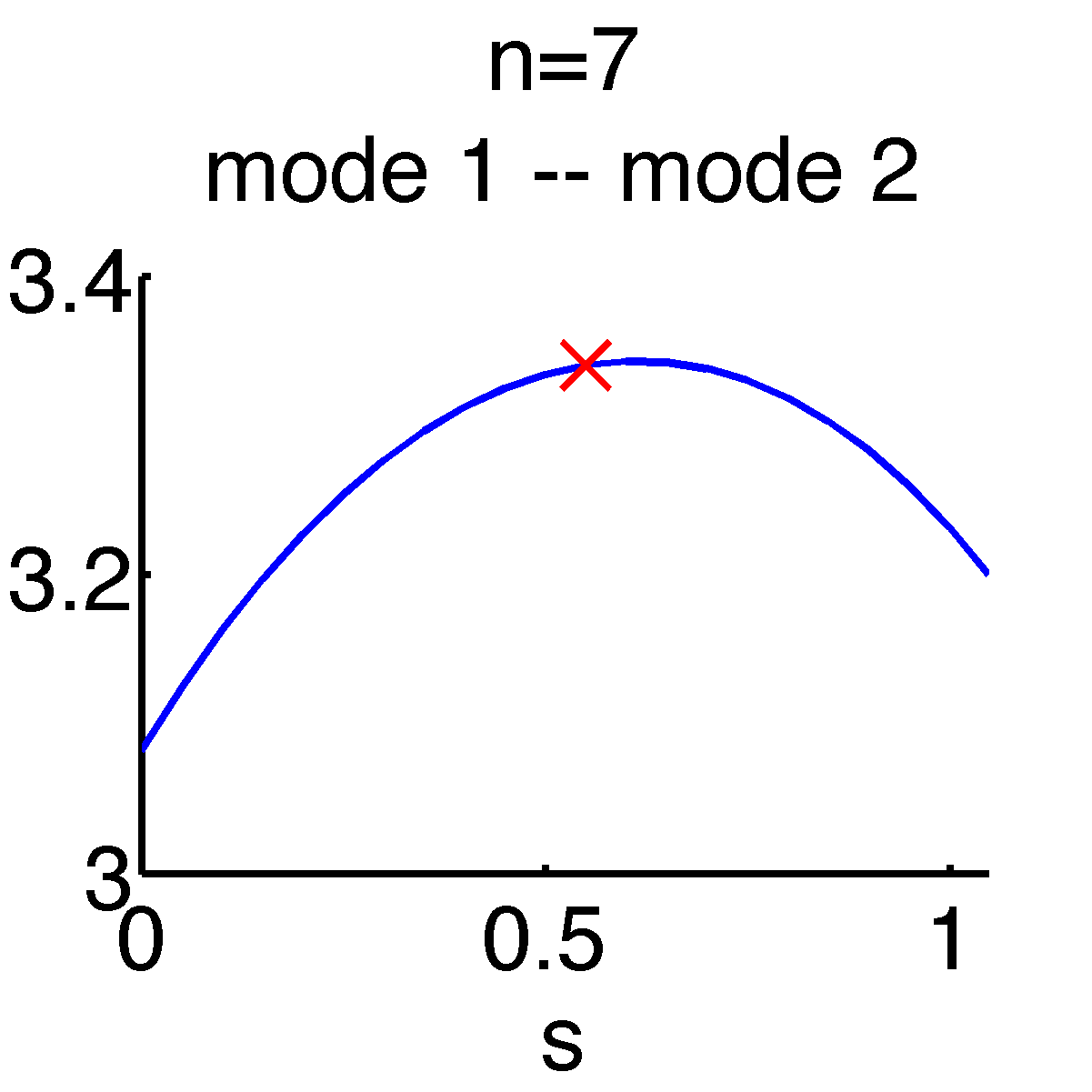}\\
\includegraphics[width=0.2\linewidth]{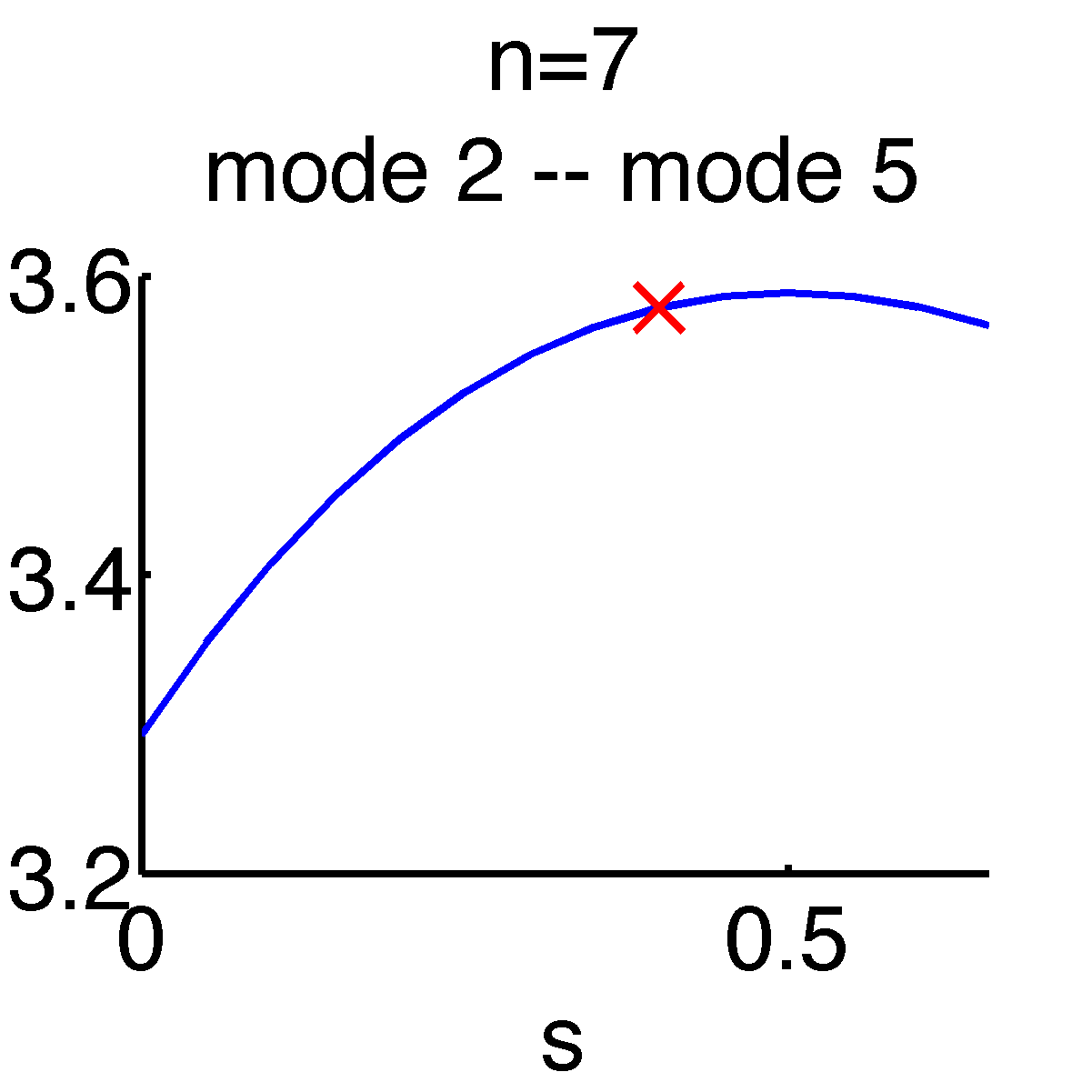}
\includegraphics[width=0.2\linewidth]{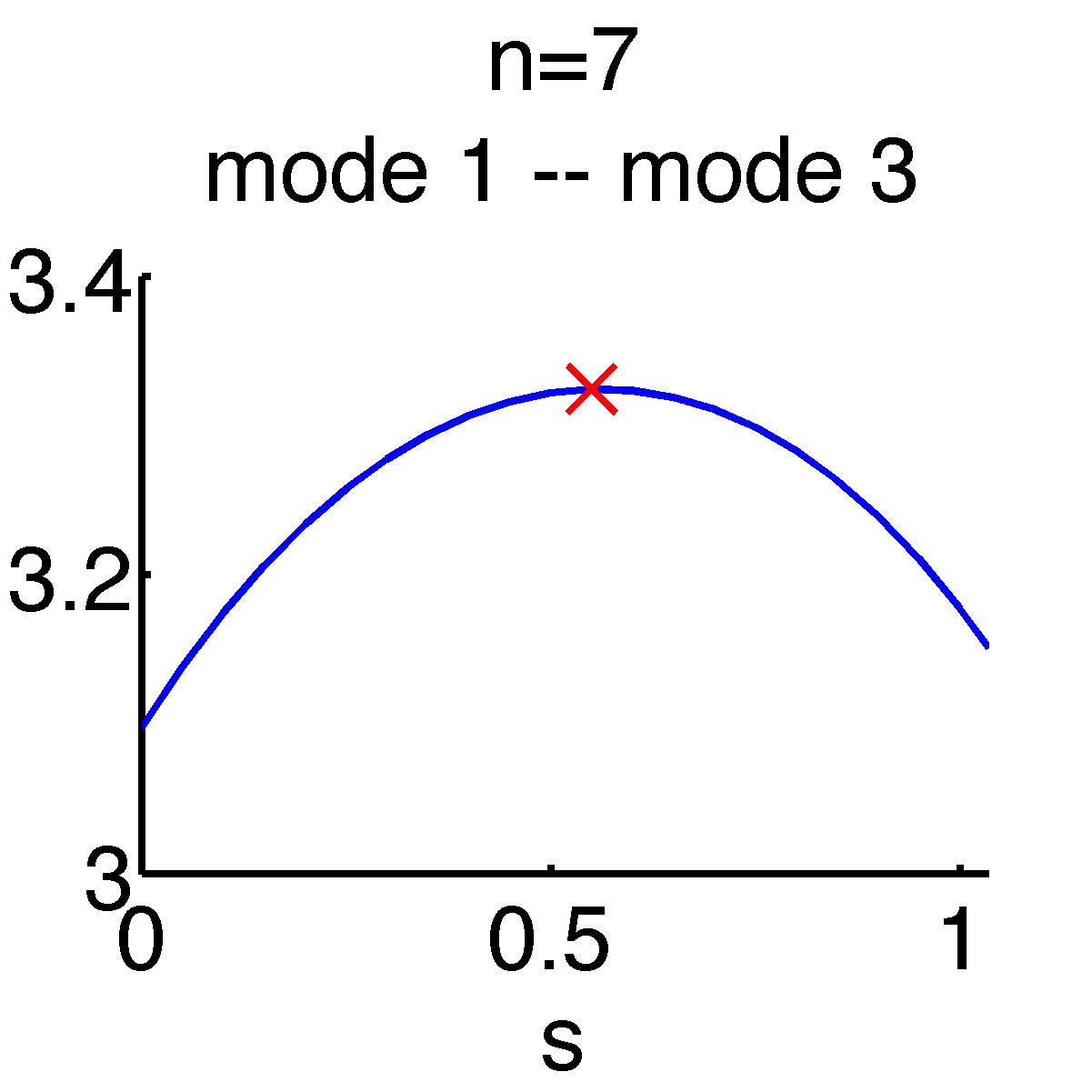}
\includegraphics[width=0.2\linewidth]{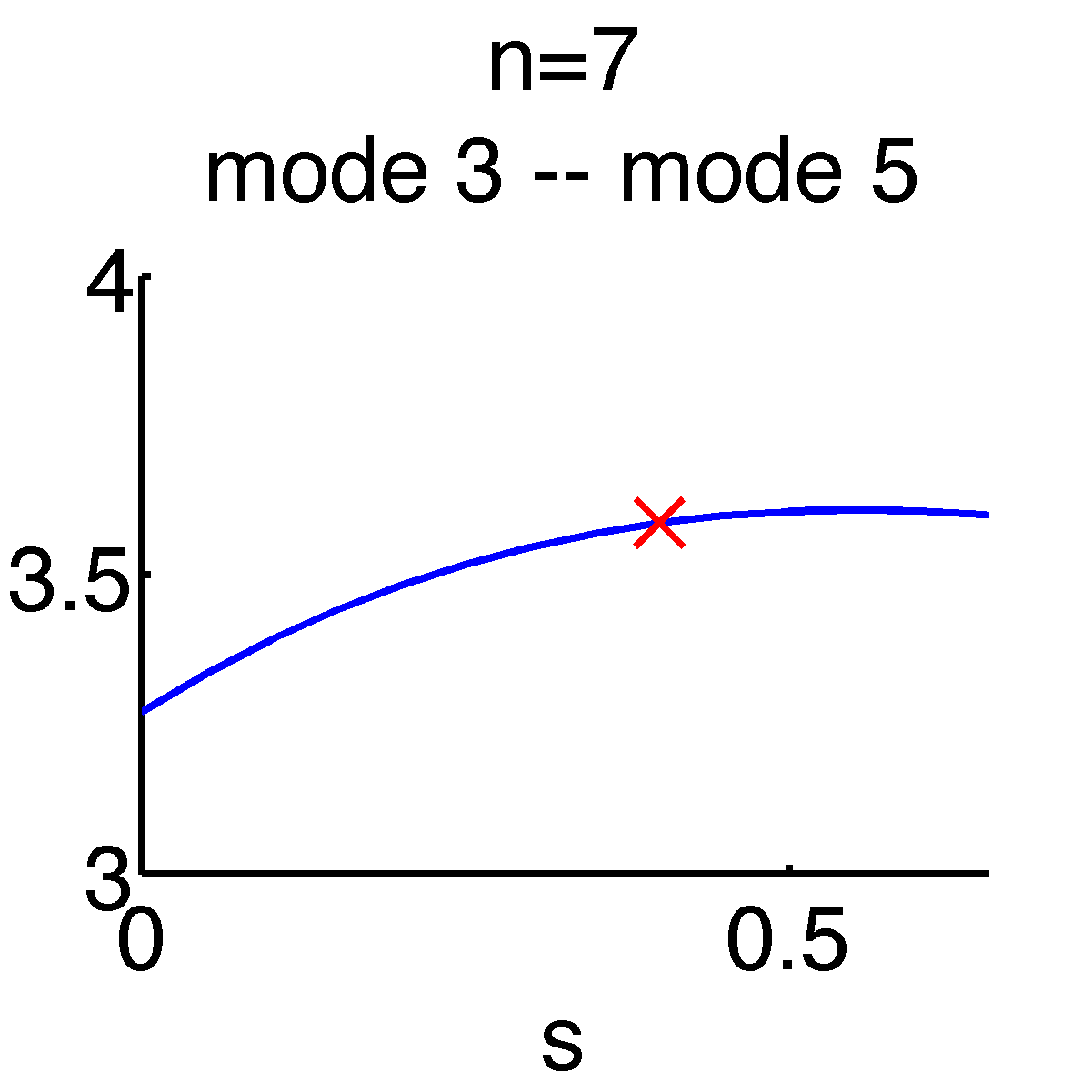}
\includegraphics[width=0.2\linewidth]{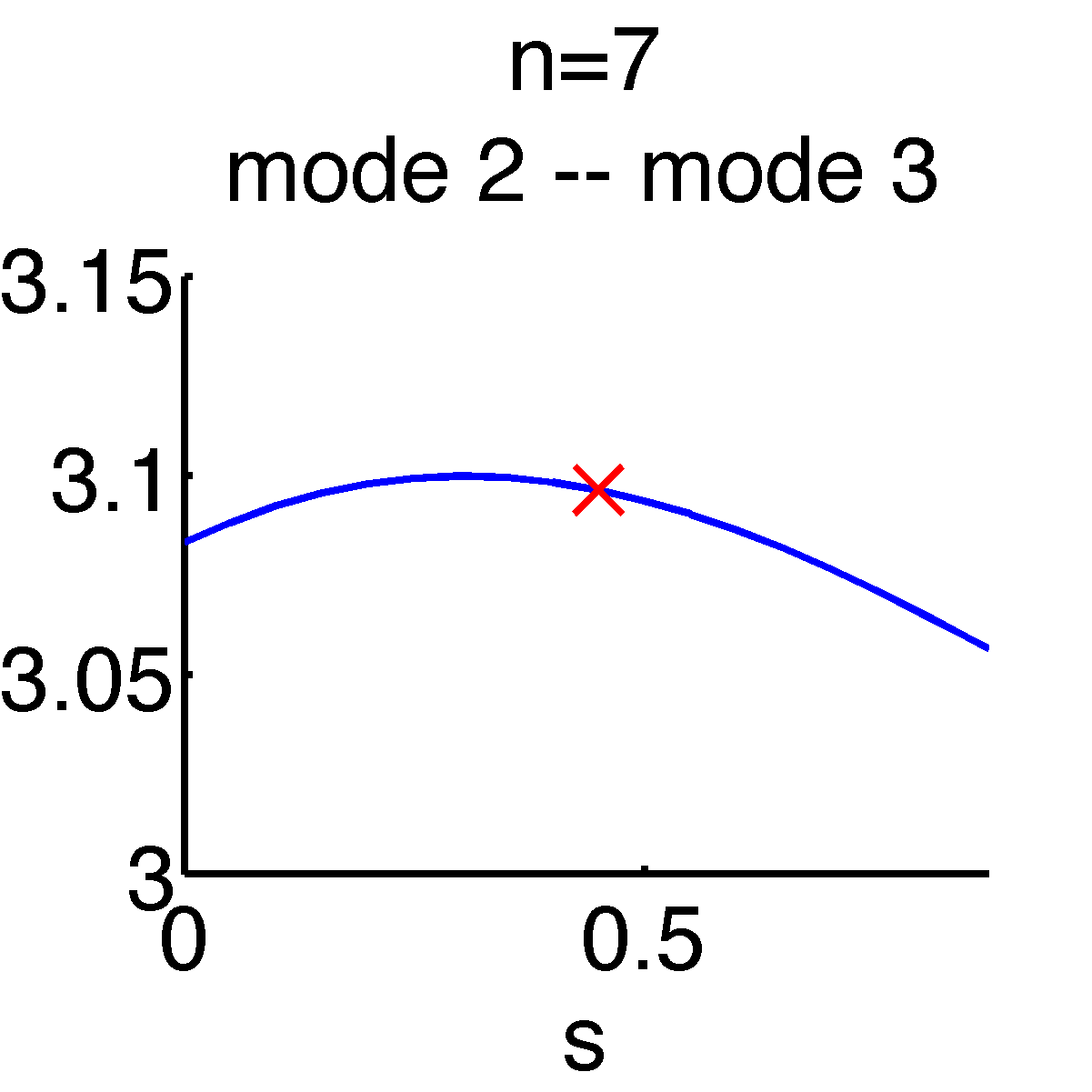}
\end{center}
\caption{Vibrational contribution to the free energy $-\log h(x)$ 
along selected 1-dimensional manifolds (red line). 
Markers indicate the transition state computed in \cite{calvo2012}
for a potential with finite width. 
 }\label{fig:floppywales}
\end{figure*}

\begin{figure*}[h]
\caption{Rigid clusters for $n=6$. Right, Mode 1 (polytetrahedron), Left, Mode 2 (octahedron).
The table below shows which bonds to break to reach each floppy mode.}
\label{fig:clustersn6}
\begin{center}
\includegraphics[width=0.35\linewidth]{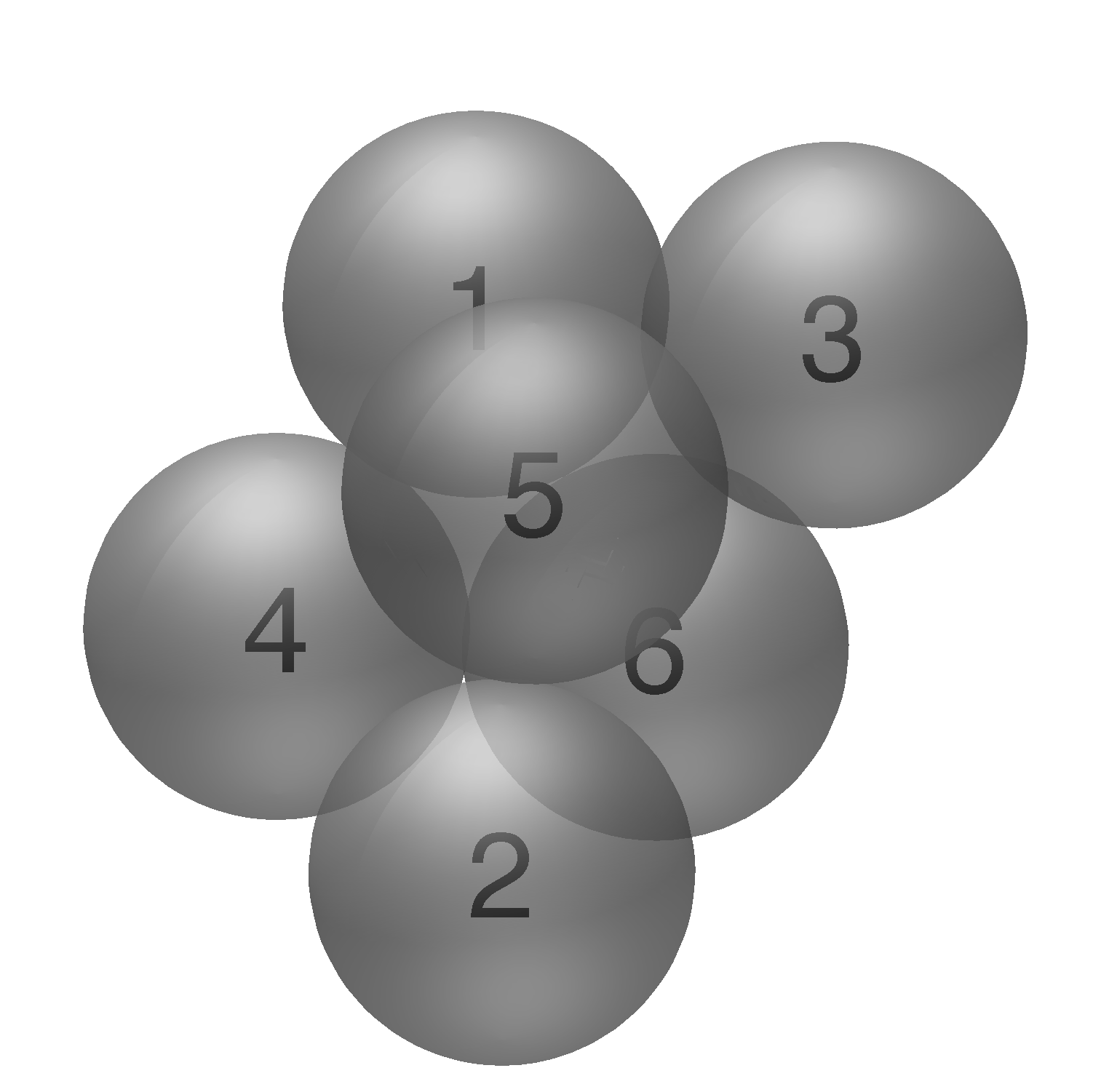}\hspace{2cm}
\includegraphics[width=0.35\linewidth]{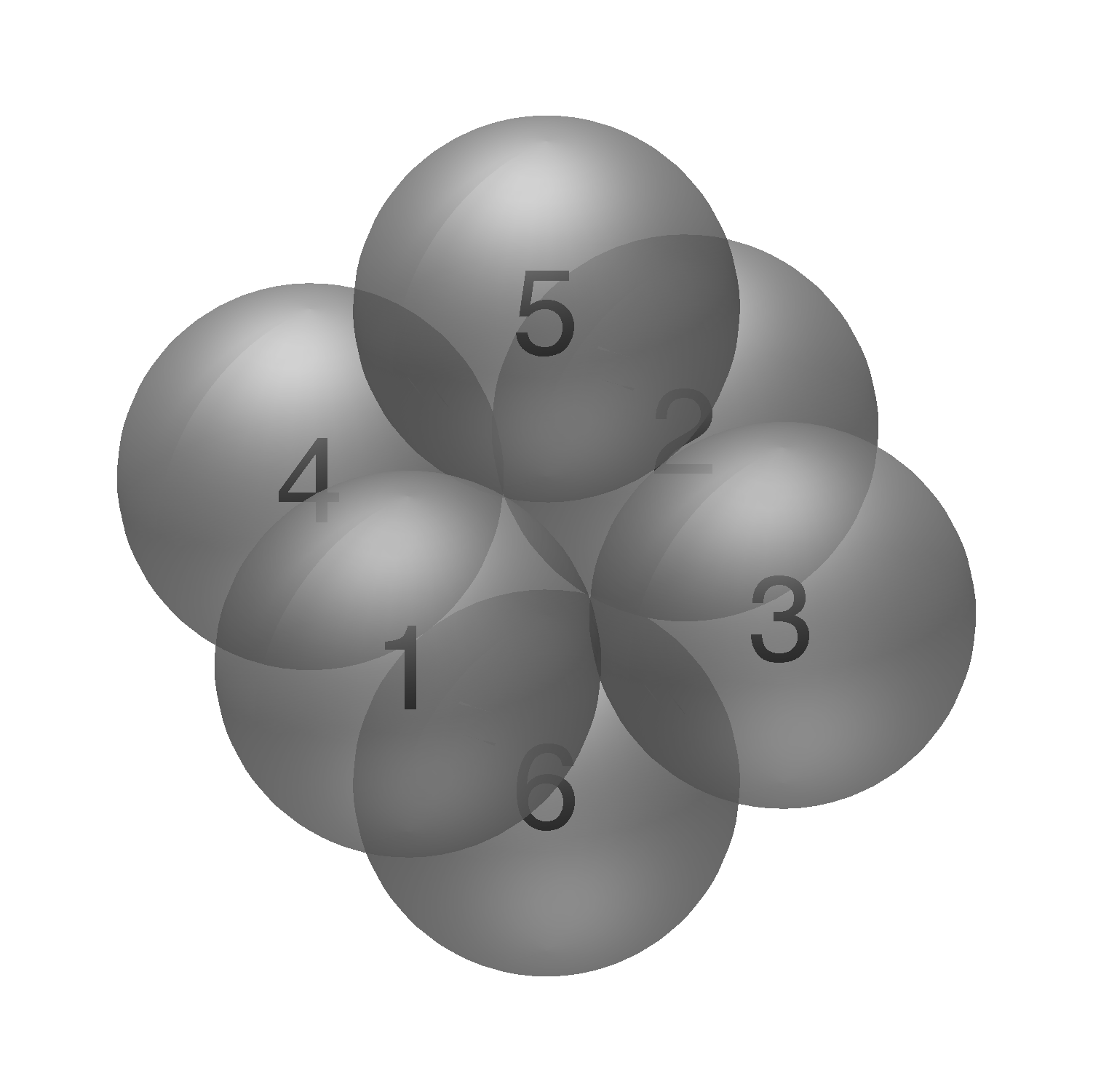}
\end{center}
\vspace{0.5cm}
\begin{tabular}{c c p{10cm}}
\hline
Mode & Starting Cluster & break bonds \\\hline
  3 &    1 & 1-3, 2-4 \\
  4 &    1 & 1-4 \\
  5 &    1 & 1-5, 4-5, 1-6, 4-6 \\
  6 &    1 & 2-5, 3-5, 2-6, 3-6 \\
  7 &    1 & 5-6 \\
 &   2 & 1-3, 2-3, 1-4, 2-4, 1-5, 2-5, 3-5, 4-5, 1-6, 2-6, 3-6, 4-6 \\
  8 &    1 & \{1-3, 1-4\}, \{1-3, 2-4\}, \{1-3, 2-4\}\\
   9 &    1 & \{1-3, 1-5\}, \{1-3, 2-5\}, \{1-3, 1-6\}, \{1-3, 2-6\}, \{2-4, 3-5\}, \{2-4, 4-5\}, \{2-4, 3-6\}, \{2-4, 4-6\}, \{2-5, 3-5\}, \{2-5, 3-5\}\\
  10 &    1 & \{1-3, 3-5\}, \{1-3, 3-6\}, \{2-4, 2-5\}, \{2-4, 2-5\}\\
  11 &    1 & \{1-3, 4-5\}, \{1-3, 4-6\}, \{2-4, 1-5\}, \{2-4, 1-6\}, \{1-5, 2-5\}, \{3-5, 4-5\}, \{1-6, 2-6\}, \{1-6, 2-6\}\\
  12 &    1 & \{1-3, 5-6\}, \{1-3, 5-6\}\\
  &   2 & \{1-3, 2-3\}, \{1-3, 1-4\}, \{2-3, 2-4\}, \{1-4, 2-4\}, \{1-5, 2-5\}, \{1-5, 1-6\}, \{2-5, 2-6\}, \{3-5, 4-5\}, \{3-5, 3-6\}, \{4-5, 4-6\}, \{1-6, 2-6\}, \{1-6, 2-6\}\\
  13 &    1 & \{1-4, 1-5\}, \{1-4, 2-5\}, \{1-4, 3-5\}, \{1-4, 4-5\}, \{1-4, 1-6\}, \{1-4, 2-6\}, \{1-4, 3-6\}, \{1-4, 4-6\}, \{1-5, 3-5\}, \{2-5, 4-5\}, \{1-6, 3-6\}, \{1-6, 3-6\}\\
  14 &    1 & \{1-4, 5-6\}, \{1-4, 5-6\}\\
  &   2 & \{1-3, 2-4\}, \{2-3, 1-4\}, \{1-5, 2-6\}, \{2-5, 1-6\}, \{3-5, 4-6\}, \{3-5, 4-6\}\\
  15 &    1 & \{1-5, 4-5\}, \{1-5, 4-5\}\\
  16 &    1 & \{1-5, 1-6\}, \{1-5, 3-6\}, \{2-5, 4-6\}, \{3-5, 1-6\}, \{4-5, 2-6\}, \{4-5, 2-6\}\\
  17 &    1 & \{1-5, 2-6\}, \{2-5, 1-6\}, \{2-5, 5-6\}, \{3-5, 4-6\}, \{3-5, 5-6\}, \{4-5, 3-6\}, \{2-6, 5-6\}, \{2-6, 5-6\}\\
  &   2 & \{1-3, 1-5\}, \{1-3, 3-5\}, \{1-3, 1-6\}, \{1-3, 3-6\}, \{2-3, 2-5\}, \{2-3, 3-5\}, \{2-3, 2-6\}, \{2-3, 3-6\}, \{1-4, 1-5\}, \{1-4, 4-5\}, \{1-4, 1-6\}, \{1-4, 4-6\}, \{2-4, 2-5\}, \{2-4, 4-5\}, \{2-4, 2-6\}, \{2-4, 4-6\}, \{1-5, 3-5\}, \{1-5, 4-5\}, \{2-5, 3-5\}, \{2-5, 4-5\}, \{1-6, 3-6\}, \{1-6, 4-6\}, \{2-6, 3-6\}, \{2-6, 3-6\}\\
  18 &    1 & \{1-5, 4-6\}, \{1-5, 5-6\}, \{4-5, 1-6\}, \{4-5, 5-6\}, \{1-6, 5-6\}, \{1-6, 5-6\}\\
  &   2 & \{1-3, 2-5\}, \{1-3, 4-5\}, \{1-3, 2-6\}, \{1-3, 4-6\}, \{2-3, 1-5\}, \{2-3, 4-5\}, \{2-3, 1-6\}, \{2-3, 4-6\}, \{1-4, 2-5\}, \{1-4, 3-5\}, \{1-4, 2-6\}, \{1-4, 3-6\}, \{2-4, 1-5\}, \{2-4, 3-5\}, \{2-4, 1-6\}, \{2-4, 3-6\}, \{1-5, 3-6\}, \{1-5, 4-6\}, \{2-5, 3-6\}, \{2-5, 4-6\}, \{3-5, 1-6\}, \{3-5, 2-6\}, \{4-5, 1-6\}, \{4-5, 1-6\}\\
  19 &    1 & \{2-5, 2-6\}, \{2-5, 2-6\}\\
  20 &    1 & \{2-5, 3-6\}, \{2-5, 3-6\}\\
 \hline\\
\end{tabular}
\end{figure*}

\section{Data}

We provide details for the floppy modes for $n=6$. Details for $n=7,8$ are available upon request. 

Table \ref{tbl:data} reports the following quantities for each mode: the volume  $S \equiv \int_{\Omega^Q_\alpha}1$\;, mean geometrical sticky parameter $\bar{h} \equiv  (\int_{\Omega^Q_\alpha}  h ) / S$, mean rotational contribution $\bar{I} \equiv  (\int_{\Omega^Q_\alpha} I ) / S$, multiplicity $n_\alpha$ (divided by a constant). Each of the integrals is over a single manifold in the set of isomorphic manifolds, and the average for rigid modes is simply the point value. For floppy modes we also report the corners, in the order they occur as one travels around the boundary.

Figure \ref{fig:clustersn6} provides a way to identify the individual modes. It shows the rigid clusters with particles numbered. The table indicates how to reach each floppy mode by starting with a given rigid structure and breaking selected bonds. In most cases there are multiple ways to reach each floppy structure and every possible bond combination that does so is listed. 

Figure \ref{fig:floppywales} plots the transition states computed in \cite{calvo2012} along each of the identified 1-dimensional floppy manifolds.

\end{document}